\documentclass[aps,prb,showpacs,twocolumn,floatfix,letterpaper,superscriptaddress]{revtex4-1}

\usepackage[intlimits,sumlimits]{amsmath}
\usepackage{amsfonts,amssymb}
\usepackage{bm}
\usepackage{graphicx}
\usepackage{cases}
\usepackage{mathtools}
\usepackage{hyperref}
\hypersetup{%
pdftitle={Tunable plasmonic reflection by bound 1D electron states in a 2D Dirac metal}, %
pdfauthor={B.-Y. Jiang et al.},%
pdfpagemode={UseNone},%
pdfstartview={FitH},%
breaklinks=true,%
citecolor=blue,%
colorlinks=true,%
linkcolor=blue,%
urlcolor=blue}

\usepackage[T1]{fontenc}
\usepackage{fourier}
\usepackage{baskervald}

\newcommand*\unit[1]{\,\mathrm{#1}} % for units in math mode

\begin{document}
\newcommand\UCSD{Department of Physics, University of California San Diego, 9500 Gilman Drive, La Jolla, California 92093}
\newcommand\UCR{Department of Physics, University of California Riverside, 900 University Avenue, Riverside, California 92521}

\title{Tunable plasmonic reflection by bound 1D electron states in a 2D Dirac metal}

\author{B.-Y.~Jiang}
\affiliation{\UCSD}
\author{G.~X.~Ni}
\affiliation{\UCSD}
\author{C.~Pan}
\affiliation{\UCR}
\author{Z.~Fei}
\affiliation{\UCSD}
\affiliation{Department of Physics, Iowa State University, 2334 Pammel Drive, Ames, Iowa 50011}
\author{B.~Cheng}
\affiliation{\UCR}
\author{C.~N.~Lau}
\affiliation{\UCR}
\author{M.~Bockrath}
\affiliation{\UCR}
\author{D.~N.~Basov}
\affiliation{\UCSD}
\affiliation{Department of Physics, Columbia University, New York, New York 10027}
\author{M.~M.~Fogler}
\affiliation{\UCSD}

\date{\today}

\begin{abstract}

We show that surface plasmons of a two-dimensional Dirac metal such as graphene can be reflected by line-like perturbations hosting one-dimensional electron states. 
The reflection originates from a strong enhancement of 
the local optical conductivity caused by optical transitions
involving these bound states.
We propose that the bound states
can be systematically created, controlled, and liquidated
by an ultranarrow electrostatic gate.
Using infrared nanoimaging,
we obtain experimental evidence
for the locally enhanced conductivity of graphene induced by a carbon nanotube gate, which supports this theoretical concept.

\end{abstract}

\maketitle

Plasmon scattering and plasmon losses in Dirac materials,
such as graphene and topological insulators, are problems of interest to both fundamental and applied research.
It is an outstanding challenge to understand
various kinds of interaction (electron-electron, electron-phonon, electron-photon, electron-disorder)
responsible for these complex phenomena~\cite{DasSarma2011ett, Kotov2012eei, Basov2014cgs, Principi2014ple, Woessner2014hcl}.
At the same time, control of plasmon scattering is critical if this class of materials is to become a new platform for nanophotonics~\cite{Jablan2009pig, Vakil2011tou, Grigorenko2012gp, deAbajo2014gpc}.

\begin{figure}[b]
\includegraphics[width=3.3 in]{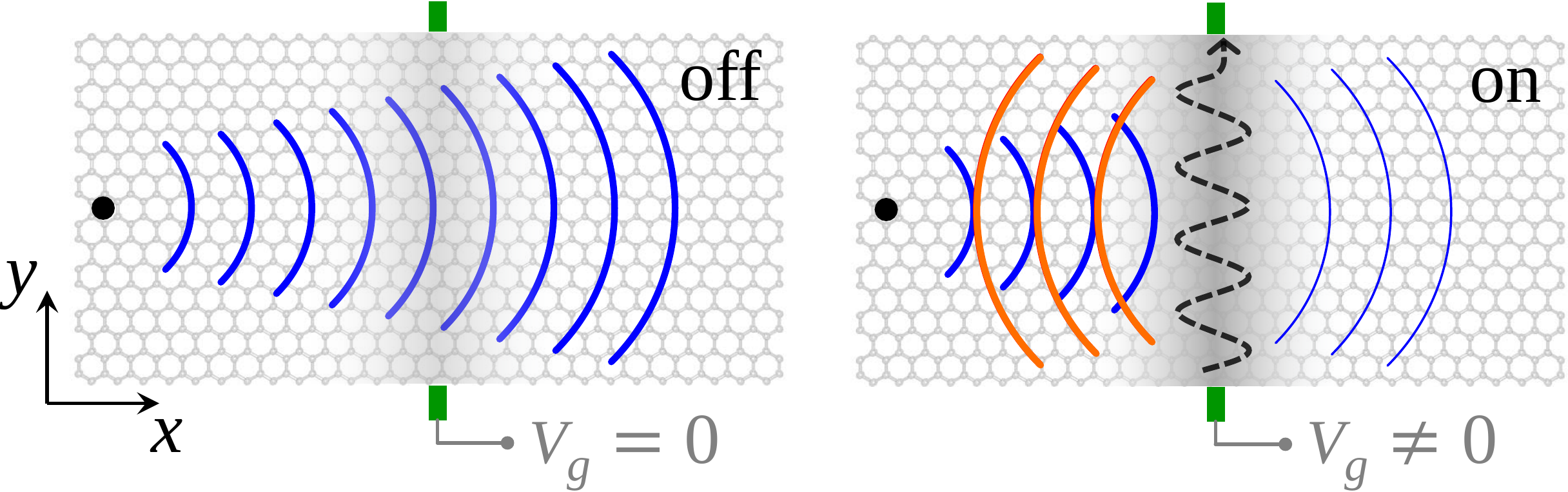}
\caption{(Color online) Schematic of an ultranarrow plasmon reflector. 
The incident plasmon (blue) can propagate freely unless
a local perturbation hosting a 1D electron state (the dashed arrow) causes it to be reflected (orange).
The bound state parameters are controlled by voltage $V_g$ of a nanotube gate (green).
}
\label{fig:switch}
\end{figure}

One source of plasmon scattering is long-range inhomogeneity
of the electron density,
which causes local fluctuations in the plasmon wavelength $\lambda_p$.
If the inhomogeneities are weak, those of size comparable to
the average $\lambda_p$ are expected to be the dominant scatterers~\cite{Kechedzhi2013pad, Garcia-Pomar2013sgp}
Surprisingly, recent experiments have revealed that one-dimensional (1D) defects of nominally atomic width can act as effective reflectors for plasmons
with wavelengths as large as a few hundred nm.
Strong plasmon reflection was observed near grain boundaries~\cite{Fei2013epp, Schnell2014soh}, topological stacking faults~\cite{Ju2015tvt}, as well as nanometer-scale wrinkles and cracks~\cite{Fei2013epp, Garcia-Pomar2013sgp} in graphene.
If this anomalous reflection is indeed an ubiquitous effect
 largely unrelated to the specific nature of a defect,
it calls for a universal explanation.  
In this Letter we attribute its origin to electron bound states commonly occurring near 1D defects.
We show that optical transitions involving the bound states
can produce strong dissipation at small distances $x$ from the defect and therefore, alter plasmon dynamics.
To support this idea we present a theoretical analysis of an exactly solvable model, which illustrates qualitative and
quantitative characteristics of the bound states and predicts how their optical response depends on the tunable parameters of a 1D potential well.
We also report an attempt to probe the predicted effects experimentally.
Our approach is to employ an ultranarrow electric gate
in the form of a carbon nanotube (CNT)
to create a precisely tunable 1D barrier in graphene.
This device enables a systematic investigation and
control of plasmon propagation, including, in principle, an
implementation of a plasmon on-off switch (Fig.~\ref{fig:switch}).
What we find is that the measured real-space
profile of the plasmon amplitude (Fig.~\ref{fig:SNOM}) cannot be accounted for
by a local change in $\lambda_p$ alone.  
Instead, the data are consistent with the
presence of an enhanced dissipation in the region next to the CNT.
The amount of this dissipation agrees in 
the order of magnitude with the power absorption due to 1D bound states in our model. 

\begin{figure}[t]
	\includegraphics[height=1.51 in]{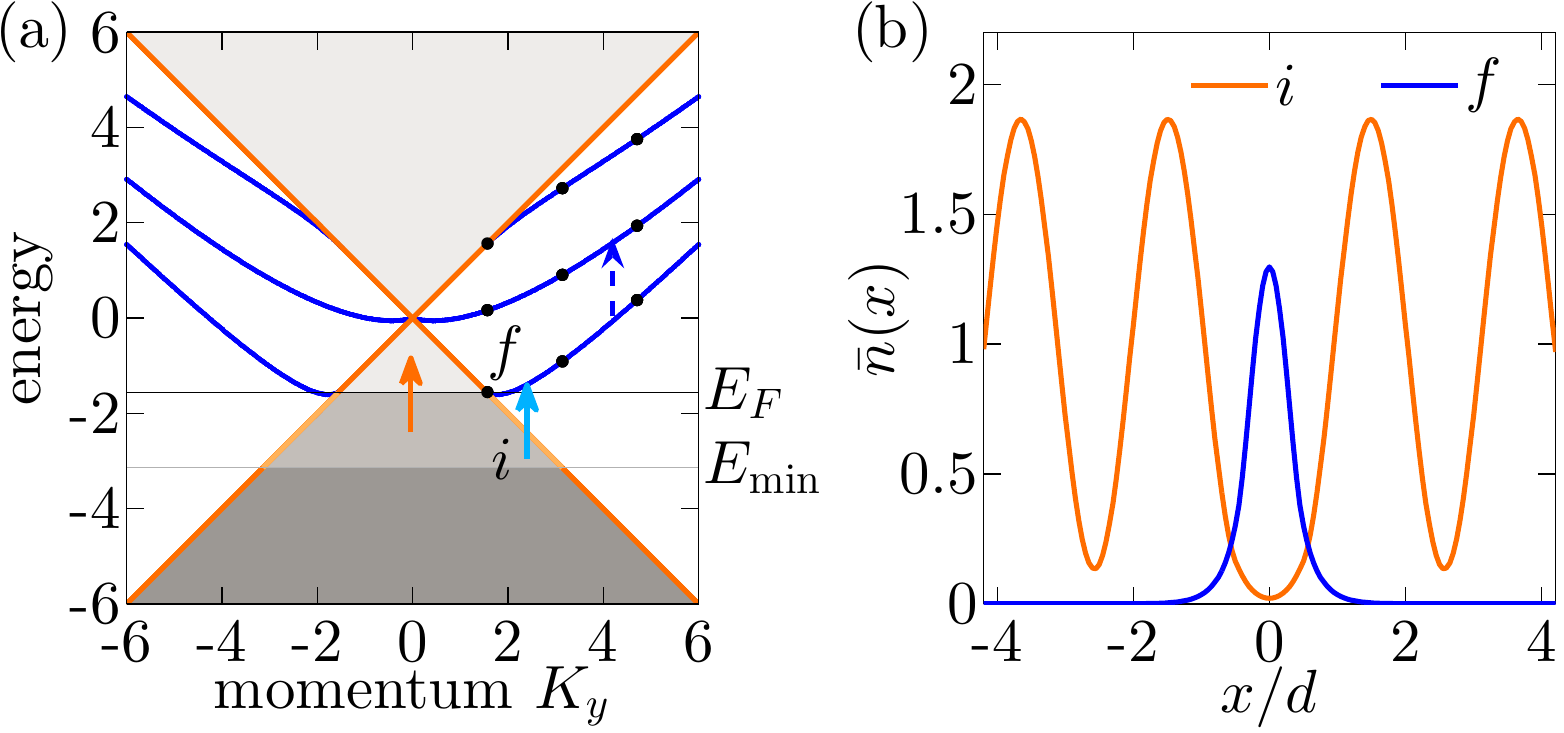}
	\caption{(Color online) (a) Dispersion of bound states for a sheet (blue) or a ribbon of width $2d$ (the black dots) for $U = 5$.
	The light gray are empty states in the continuum. The dark and medium gray are occupied states in the continuum. The last of these, with $E$ between 
	$E_F = \mu d/ \hbar v_F$ and $E_\mathrm{min} = E_F - \omega d/v_F$, enable optical transitions (the arrows) of frequency $\omega$. 
	Transitions between bound states (the dashed arrow) can occur 
	for some $E_F$, e.g., $E_F = 0$ at which the state
	$i$ is filled and the state $f$ is empty.
	(b) The density distribution $\bar{n}=|\Psi|^2$ of the two states $i$ and $f$ for the transition indicated by the cyan arrow in (a).
	The state $i$ (blue) is localized in the well, while 
	the state $f$ (orange) is extended.
	Parameters: $K_y=2.5$, $\omega d/v_F = \pi/2$.
	}
	\label{fig:bound_state}
\end{figure}

\textit{Model.}---We assume that the graphene quasiparticles can be described by a 2D Dirac Hamiltonian $H = \hbar v_F\left(\sigma_z k_x+\sigma_y k_y\right) + v(x)$,
where $\sigma_y$, $\sigma_z$ are the Pauli matrices and $v(x)$ is
the total (screened) potential induced by the 1D gate.
For simplicity, we assume that $v(x)$ is a square well of width $d$ and depth $u$ although more realistic potentials~\cite{Kennedy2002wsp, Hartmann2014qes, Candemir2013mde, Villalba2003trs}
can also be considered.
In the present case the eigenfunctions $\Psi$ are combinations of plane waves and/or exponentials that have to be matched at $x = \pm d / 2$,
see Appendix~\ref{sec:square_well}.
The electron momentum $k_y$ along the
perturbation (in the $y$-direction) is
conserved, so that the gapless 2D Dirac spectrum is effectively replaced by a 1D one with a gap $\Delta = |\hbar v_F k_y|$.
Within the gap electron states localized at the well exist [Fig.~\ref{fig:bound_state}(b)].
The energies $\varepsilon_n(k_y)$ of these bound states, where $n = 1, 2, \ldots$, are the solutions of the transcendental equation~\cite{Beenakker2009qgh}
\begin{equation}
	\frac{\tan\sqrt{(E + U)^2 - K_y^2}}{\sqrt{(E + U)^2 - K_y^2}}
	= \frac{i \sqrt{E^2 - K_y^2}}{K_y^2 - E (E + U)} \,.
	\label{eqn:loc_dispersion}
\end{equation}
Here $E = \varepsilon_n d / (\hbar v_F)$ is the dimensionless energy and
\begin{equation}
	K_y =  k_y d\,,
	\quad
	U =  u d / (\hbar v_F)\,,
\end{equation}
are, respectively, the dimensionless $y$-momentum and the well depth.
The dispersions of the three lowest bound states for $U = 5$ are shown in Fig.~\ref{fig:bound_state}(a).

\begin{figure}[b]
\includegraphics[width=3.2in]{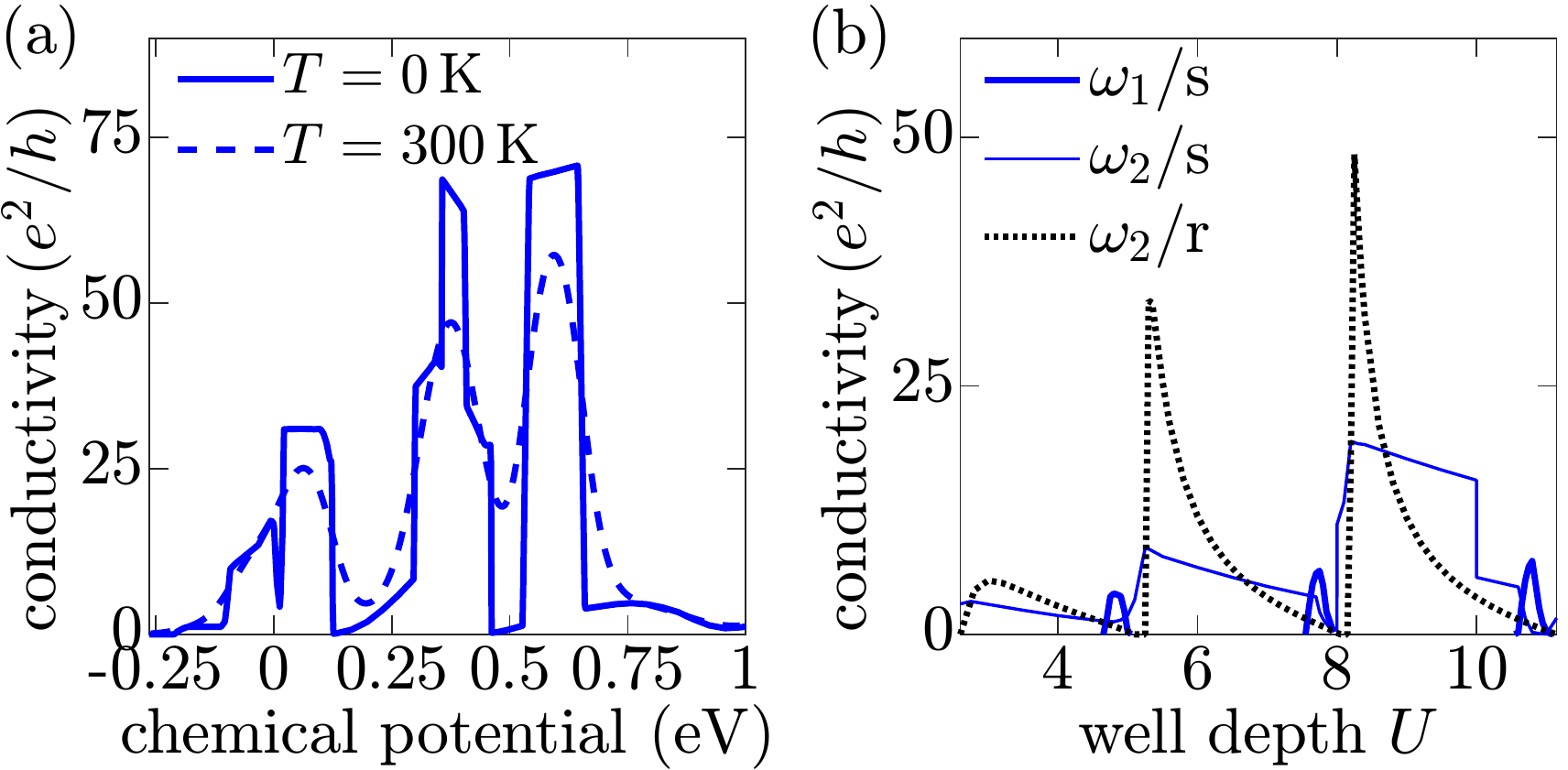}
\caption{(Color online)
(a) Integrated conductivity $\bar{\sigma}$ of a graphene sheet 
at $\omega = 830\unit{cm^{-1}}$.
The sharp changes are caused by
blocking/unblocking of the transitions involving bound states
as a result of changing occupations of the levels as a function of the graphene chemical potential $\mu$.
For example, the plateau at $0.02 < \mu(\mathrm{eV}) < 0.12$ is due to the (blue) dashed-line transition in Fig.~\ref{fig:bound_state}(a). 
(b) Integrated conductivity $\bar{\sigma}$ of a sheet (s) and a ribbon (r) at $T = 0$ and $K_F = -\pi/2$.
Sharp changes at $U = 8$ and $10$ for $\omega = \omega_2$ arise from a transition between bound states. 
Parameters: $d = 10\unit{nm}$, $\omega_1 = 83\unit{cm^{-1}}$, $\omega_2 = 830\unit{cm^{-1}}$.
}
\label{fig:optical_signal}
\end{figure}

\begin{figure}
\includegraphics[width=3.4 in]{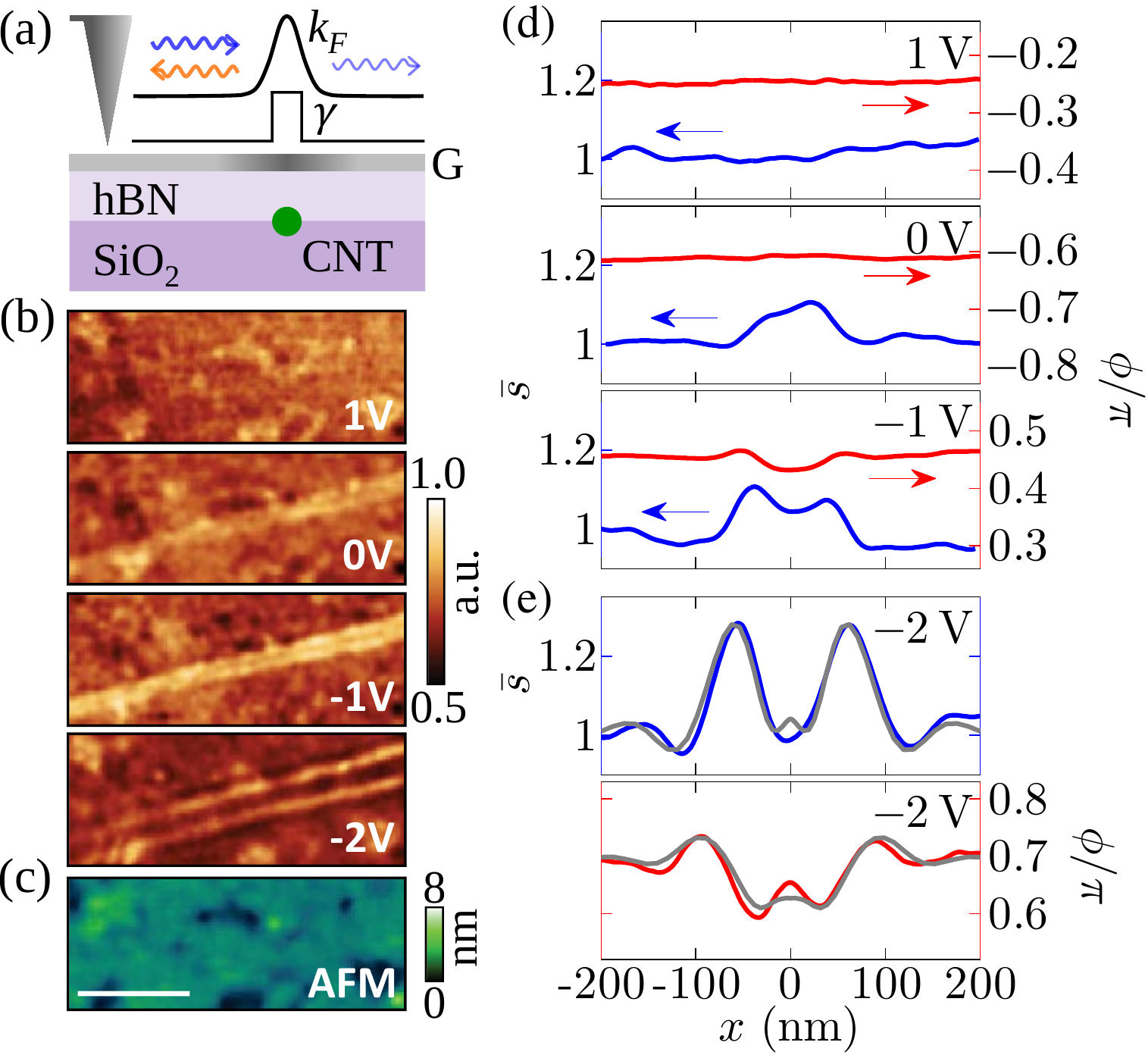}
\caption{(Color online) Measurement of the conductivity $\bar{\sigma}$ by the s-SNOM.
(a) A schematic showing
graphene (variable intensity gray) gated by a CNT (green) separated from it by a thin hBN layer.
The induced perturbation is parameterized by spatially varying $k_F$ and $\gamma$.
In the experiment, the AFM tip (triangle) is polarized by a focused infrared beam (not shown),
which enables it to launch a plasmon (blue).
The reflected plasmon (orange)
causes an additional tip polarization,
resulting in a modified optical signal backscattered by the tip and detected in the far field. 
(b) The s-SNOM amplitude images of the region next to the CNT for $V_g = +1...-2\unit{V}$ and $\omega = 890\unit{cm}^{-1}$.
The twin fringes (bright lines)
intensify and separate as $|V_g|$ increases.
(c) The AFM topography image of the same region. Scale bar: $1\,\mu\mathrm{m}$.
(d)-(e) The s-SNOM amplitude ($\bar{s}$) and phase ($\phi$) 
along the line perpendicular to the CNT;
$\bar{s}$ is normalized to $x = -200\unit{nm}$ point.
The
best theoretical fits (gray) for $V_g = -2\unit{V}$ are included in (e).
}
\label{fig:SNOM}
\end{figure}

The response of the system to
an optical excitation of frequency $\omega$ polarized in the $x$-direction
is described by an effective conductivity $\sigma(x)$ given by the Kubo formula~\cite{supp_mat},
which determines the local current density $j_x(x) =  E_x \sigma(x)$ in the approximation that the total electric field $E_x$ due to the optical excitation is uniform.
Below we focus on the real part of $\sigma(x)$, which determines local power dissipation.
We assume that graphene is doped and consider only frequencies $\hbar\omega < 2 |\varepsilon_F|$,
for which
the optical conductivity of an infinite graphene sheet vanishes
(if we neglect disorder, many-body scattering, and thermal broadening~\cite{Basov2014cgs}).
This implies that in the absence of the perturbation, $U = 0$,
we must have $\mathrm{Re}\,\sigma(x) = 0$ at all $x$.
On the other hand, when the potential well is present,
a finite $\mathrm{Re}\,\sigma(x)$ exists.
There are two types of relevant optical transitions:
those that involve the bound states [as either the initial $i$ or the final $f$ states,
Fig.~\ref{fig:bound_state}(a)] and those that do not.
The contribution of the former to $\mathrm{Re}\,\sigma(x)$ is maximized
near the potential well and decays exponentially at $|x| > d / 2$
due to the localized nature of the bound states.
The contribution of the latter is small, 
oscillating, and decaying algebraically with $x$~\cite{supp_mat}.
Resolving the detailed real-space features of $\sigma(x)$ in an optical experiment is challenging (see below). A more practical observable is the normalized integrated conductivity:
\begin{equation}
\bar{\sigma} \equiv \frac{1}{d} \int\limits_{-\infty}^{\infty}
d x\,\mathrm{Re}\, \sigma(x)\,.
\label{eqn:sigma_bar_def}
\end{equation}
According to our simulations,
transitions that involve the bound states
give the dominant contribution to $\bar{\sigma}$.
In particular, bound-to-bound state transitions produce numerically large values of $\bar{\sigma}$ expressed in units of $e^2 / h$.
Such transitions are possible at discrete $k_y$ where the energy difference between the states of the same momentum matches $\hbar \omega$ provided the lower (higher) state is occupied (empty). 
If the chemical potential $\mu$ is gradually increased, e.g., by electrostatic gating, the state occupations would change, leading to
either blocking or unblocking of these transitions.
Accordingly, $\bar{\sigma}$ would either sharply drop or jump, see
Fig.~\ref{fig:optical_signal}(a).
These changes persist, albeit blurred, at
finite temperatures,
see the dashed curve in Fig.~\ref{fig:optical_signal}(a).

Sharp drops in $\bar{\sigma}$ also occur when the bound states merge with the continuum and get liquidated (become extended).
The drop is abrupt if the optical transitions probe a single $k_y$ or a narrow range of $k_y$.
In principle, this situation can be realized in a graphene ribbon
running perpendicular to the linelike perturbation.
In such a ribbon the allowed $k_y = m \pi / W + \mathrm{const}$ are discrete,
as shown schematically by the dots in Fig.~\ref{fig:bound_state}(b).
The coupling to a single bound state
can be achieved under the condition $\pi / W > \omega / v_F$, i.e.,
by using a ribbon of a narrow width $W$ or the excitation of a low frequency $\omega$.
In Fig.~\ref{fig:optical_signal}(b) we show three numerically calculated traces of $\bar{\sigma}$ as a function of the well depth $U$ for a fixed dimensionless chemical potential $E_F = \mu d / (\hbar v_F) = -\pi/2$.
The first trace is computed for a ribbon of width $W = 2d$
probed at the excitation energy $\hbar\omega = |\mu|$.
It exhibits pronounced oscillations of $\bar{\sigma}$.
In particular, $\bar{\sigma}$ drops to zero when a bound state merges with the continuum.
The other two traces correspond to a 2D graphene sheet.
Although the sharp drops become blurred,
they remain pronounced at a low excitation energy 
$\hbar\omega_1 = |\mu| / 10$ and
still evident at $\hbar\omega_2 = |\mu|$.

The enhanced local optical conductivity around the 1D gates described above causes plasmons to be strongly reflected.
According to the first-order perturbation theory~\cite{Fei2013epp, Garcia-Pomar2013sgp, supp_mat},
the reflection coefficient $r$ of a normally incident plasmon wave is
\begin{equation}
	r_1 \simeq \frac{2\pi i}{\lambda_p} \int\limits_{-\infty}^{\infty}dx \left[\frac{\sigma(x)}{\sigma_\infty}-1\right]\,.
\end{equation}
For arbitrary perturbations, we can use the approximation
$|r| \approx \min (\,|r_1|, 1)$. 
Using the results of Fig.~\ref{fig:optical_signal}(b) we estimate $|r| \approx 0.3$ at the chemical potential of $0.25\unit{eV}$ where the predicted $\bar{\sigma}\approx 5{e^2/h}$.
This roughly corresponds to the regime probed by our experiments (see below).
At the chemical potential of $0.3\unit{eV}$ where the calculated local conductivity is much larger, $\bar{\sigma}\approx 40 {e^2/h}$,
the reflection coefficient should approach unity, realizing the ``reflector on'' state in Fig.~\ref{fig:switch}.

\textit{Experiment and analysis.}---To investigate the described above phenomena experimentally we fabricated a nanodevice that contained (bottom to top)
a Si/SiO$_2$ substrate,
a $10 \unit{nm}$-thick layer of hexagonal boron nitride (hBN),
and a mechanically exfoliated graphene flake.
A metallic single-wall CNT was placed between hBN and SiO$_2$.
The local charge density of graphene was tunable by the voltage $V_g$ applied between the CNT and graphene.
The average carrier density in graphene $|n| \sim 5\times 10^{12}\unit{cm}^{-2}$ was produced by uncontrolled ambient dopants (acceptors)~\cite{Ni2015pgm}.
To infer the local optical conductivity $\sigma(x)$ we used
scattering-type scanning near-field optical microscopy (s-SNOM)~\cite{Keilmann2009, Atkin2012n, Basov2014cgs},
see Fig.~\ref{fig:SNOM}(a).
The s-SNOM utilizes a tip of an atomic force microscope (AFM)
with a radus $25\unit{nm}$ as an optical antenna that couples incident infrared light to graphene plasmons.
The backscattered light is analyzed to extract the amplitude $\bar{s}$ and the phase $\phi$ of the genuine near-field signal, Fig.~\ref{fig:SNOM}(b,d,e).
Crudely speaking, this signal is
proportional to the electric field inside the tip-sample nanogap.
The variation of this field with the tip position is caused by
the standing-wave patterns of
surface plasmons~\cite{Fei2012gtg, Chen2012oni}.
These standing waves are due to the interference of
the plasmon waves launched by the tip with
the waves reflected by the charge inhomogeneity induced by the CNT.
The spacing of the interference fringes
is equal to one half of the plasmon wavelength $\lambda_p$.
The latter is given by $\lambda_p = \mathrm{Re}\, (2\pi / q_p)$,
where $q_p(x) = {i \kappa \omega}/{2 \pi \sigma(x)}$
is the complex plasmon momentum and
$\kappa$ is the average permittivity of the media surrounding graphene~\cite{Basov2014cgs}. Therefore, s-SNOM images combined with the formula for $q_p$ give a direct estimate of $\mathrm{Im}\, \sigma(x)$.
The extraction of $\mathrm{Re}\, \sigma(x)$
requires an electromagnetic simulation of the coupled tip-graphene system,
which was done using the numerical algorithm developed previously~\cite{Fei2012gtg, Fei2013epp, supp_mat}.
To facilitate connection with that previous work,
we parametrized the conductivity via
\begin{equation}
\sigma(x) = \frac{e^2 v_F}{\pi \hbar\omega}\,
\frac{i k_F(x)}{1 + i \gamma(x)}\,,
\label{eqn:sigma_Drude}
\end{equation}
which was modelled after the long-wavelength Drude (intraband) conductivity of graphene~\cite{Basov2014cgs} with Fermi momentum $k_F$ and dimensionless damping factor $\gamma$.
The goal of the data analysis was to determine the profiles of $k_F(x)$ and $\gamma(x)$ that yield the best fit to the s-SNOM data. 
In this parametrization, the presence of the bound states should increase the local damping,
so the signature we were looking for was the enhanced value of $\gamma(x)$.

Our experimental data are presented in Fig.~\ref{fig:SNOM}.
The AFM topography image, Fig.~\ref{fig:SNOM}(c), 
shows that the CNT does not produce any visible topographic features.
However, in the near-field signal, up to two pairs of intereference fringes appear on each side of the CNT [the bright lines in Fig.~\ref{fig:SNOM}(b)].
Similar twin fringes have been observed in prior s-SNOM imaging~\cite{Fei2013epp, Schnell2014soh, Ju2015tvt, Ni2015pgm} of linear defects in graphene.
Importantly, the intensity and spacing of the fringes
we observe here evolve with the CNT voltage $V_g$,
which attests to their electronic (specifically, plasmonic) origin.

In addition to the controlled perturbation induced by the CNT,
graphene contains uncontrolled ones due to random defects.
To reduce the random noise caused by those,
we averaged the near-field signal over a large number of
linear traces taken perpendicular to the CNT.
Thus obtained line profiles of both the amplitude $\bar{s}$ and the phase $\phi$
are plotted in Fig.~\ref{fig:SNOM}(d) and (e).
We focus on the $V_g = -2\unit{V}$ trace, which shows the strongest modulation.
The accurate determination of functions $k_F(x)$ and $\gamma(x)$
is impacted by the s-SNOM resolution limit $\sim 20\unit{nm}$.
In our fitting we assumed that $k_F(x)$
is given by the perfect screening model,
$k_F^2(x) = [k_F^2(0) d^2 +  k_F^2(\infty) x^2] / (d^2 + x^2)$,
which should be a good approximation for high doping~\cite{Jiang2015erg}.
The adjustable parameters are $k_F(0)$ and $k_F(\infty)$.
For $\gamma(x)$ we considered trial functions in the form of a peak (dip) at $x = 0$, with adjustable width and height (depth),
as sketched in Fig.~\ref{fig:SNOM}(a).
The trial $k_F(x)$ and $\gamma(x)$ were fed
as an input to the electromagnetic solver described previously~\cite{Fei2012gtg, Fei2013epp}.
As detailed in Appendix~\ref{sec:fitting},
a good agreement
with the observed form of the twin fringes requires
a strong peak in $\gamma(x)$ near the CNT.
The shape of the fringes was found to depend primarily on the integral of $\gamma(x) - \gamma(\infty)$,
so in the end we modeled $\gamma(x)$ by a box-like discontinuity with a central region of a fixed width $13.5\unit{nm}$ and
two adjustable parameters $\gamma(0)$, $\gamma(\infty)$.
The best fits
[the gray curves in Fig.~\ref{fig:SNOM}(e)] to the 
$V_g = -2\unit{V}$ s-SNOM data were obtained using $\gamma(0)=1.65$.

To establish a rough correspondence between the profiles of Fig.~\ref{fig:SNOM}(e) and the square-well model we take $d$ to be the thickness of the hBN spacer $d = 10\unit{nm}$
and $U$ to give the same integrated weight $\int v(x) d x \equiv u d = \hbar v_F U
 = \hbar v_F \int [k_F(x) - k_F(\infty)] d x$.
This prescription implies $E_F = 4$, $U = 13$, and
$\bar{\sigma} = 3.5\,e^2/h$
for $\omega = 890\unit{cm}^{-1} = 1.7 v_F / d$~\cite{supp_mat}.
The square-well model in Fig.~\ref{fig:optical_signal}
yields a comparable optical conductivity $\bar{\sigma} = 4.7\,e^2/h$
although for a smaller $U = 5$.
Given a number of simplifying assumptions we have made in the modelling,
this level of agreement seems adequate.

\begin{figure}
\includegraphics[height=2.1 in]{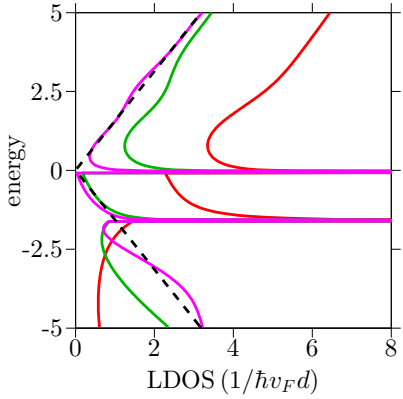}
\caption{(Color online)
The LDOS as a function of the dimensionless energy $E$ for the $U = 5$ square-well model at the
three fixed distances from the CNT: $x/d = 0$ (red), $0.6$ (green), and $1.0$ (violet).
The dashed line shows the LDOS of unperturbed graphene.
}
\label{fig:LDOS}
\end{figure}

\textit{Summary and future directions.}---In this Letter, we proposed a model for the anomalous plasmon reflection by ultranarrow electron boundaries in graphene.
We validated this concept in experiments with electrostatically tunable line-like perturbations.
One broad implication of our work is that nanoimaging of collective modes can reveal nontrivial electron properties, in this case, 1D bound states.
Recent experiments have demonstrated that this technique is not limited to plasmons or graphene or 2D systems~\cite{Dai2014tpp, Shi2015oll, Giles2016iai, Fei2016noi}.
We hope that our work stimulates even wider use of this novel spectroscopic tool.

A particularly intriguing future direction is to complement s-SNOM with
scanned probe techniques other than AFM topography.
For example, scanning tunneling microscopy, which has a superior spatial resolution,
can be used to measure the local electron density of states (LDOS).
For the particular model system studied here,
the features exhibited by the LDOS should be quite striking, see 
Fig.~\ref{fig:LDOS} and Appendix~\ref{sec:LDOS}.
The origin of these features can be understood by examining the dispersions
in Fig.~\ref{fig:bound_state}(a).
Within the selected energy interval there is the total of three bound states. 
The topmost one has a monotonic dispersion; the other two have energy minima at which the LDOS
has van Hove singularities (diverges),
see Fig.~\ref{fig:LDOS}.
The strength of these singularities decreases exponentially with $x$ because these bound states are localized near the well.
At large $x$, the LDOS displays the $\mathrm{V}$-shaped energy dependence characteristic of uniform graphene~\cite{Basov2014cgs}.
We anticipate that the combination of optical and tunneling nanoimaging and nanospectroscopy could provide a refined information about the local electronic structure.
One example of a possible application of this knowledge is the design of optimized plasmon switches
(Fig.~\ref{fig:switch}) for Dirac-material-based nanoplasmonics.

We acknowledge support by the ONR under Grant N00014-13-0464 and
by the NSF under Grant ECCS-1509958 (M.B.).

\appendix
%%%%%%%%%%%%%%%%%%%%%%%%%%%%%%%%%%%%%%%%%%%%%%%%%%%%%%%%%%%%%%%%%%%%%%%%%%%
\section{Local optical conductivity of a nonuniform graphene}
\label{sec:square_well}

%The simple model of the optical conductivity [Eq.~\eqref{eqn:graphene_conductivity}] fails if the perturbation is very sharp.

In this section we describe the details of our 1D square well model, including the analytical expressions for the wavefunctions and the calculation of the optical conductivity in the vicinity the well.
We start from the 2D Dirac Hamiltonian for the quasiparticles, 
%%%
\begin{equation}
H=\hbar v_F(\sigma_z k_x+\sigma_y k_y)+v(x)\,,
\end{equation}
%%%
where $\sigma_i$ are the Pauli matrices acting on the sublattice pseudospin.
The potential $v(x)$ is taken to be a square well,
%%%
\begin{equation}
v(x)=
\begin{cases}
-u, &|x|<d/2\,,
\\
\phantom{-}0, &|x|>d/2\,.
\end{cases}
\end{equation}
Without loss of generality, we take $u$ to be positive.
As the system is invariant in the $y$-direction, $k_y$ is conserved,
and so our problem is effectively one-dimensional.
We use the following terminology: region I is the part of the system to the left of the well ($ x< -d/2$); region II is the strip containing the well ($|x|<d/2$); region III is to the right of the well ($x>d/2$).
For an eigenstate to have the same energy across all three regions,
the magnitude $k$ of its momentum must obey the following relations
\begin{equation}
k_\mathrm{I}=k_\mathrm{III}=k_\mathrm{II}-\dfrac{u}{\hbar v_F}\,,\quad k_l^2=k_{xl}^2+k_y^2\,,
\end{equation}
with $l=\mathrm{I}$, II, or III.
When $k_l>0$, the wavefunction belongs in the conduction band and has the form 
\begin{equation}
\Psi_c(\textbf{r})\propto
\psi_c(\theta_l)
e^{ik_{xl} x+ik_y y}\,,
\quad
\psi_c=
\begin{bmatrix*}[r]
i\cos(\theta_l/2)
\\
-\sin(\theta_l/2)
\end{bmatrix*}\,;
\end{equation}
when $k_l<0$, the wavefunction belongs to the valence band and is given by
\begin{equation}
\Psi_v(\textbf{r})\propto
\psi_v(\theta_l)
e^{ik_{xl} x+ik_y y}\,,
\quad
\psi_v=
\begin{bmatrix*}[r]
i\sin(\theta_l/2)
\\
\cos(\theta_l/2)
\end{bmatrix*}\,.
\end{equation}
The angle $\theta_l=\arctan(k_y/k_{xl})$ defines the direction of the momentum.
In the following we use the notation $k$ for $k_\mathrm{I}$ and $\bar{k}$ for $k_\mathrm{II}$ and similarly for all other quantities.

\begin{figure*}[t]
\includegraphics[width=2.3 in]{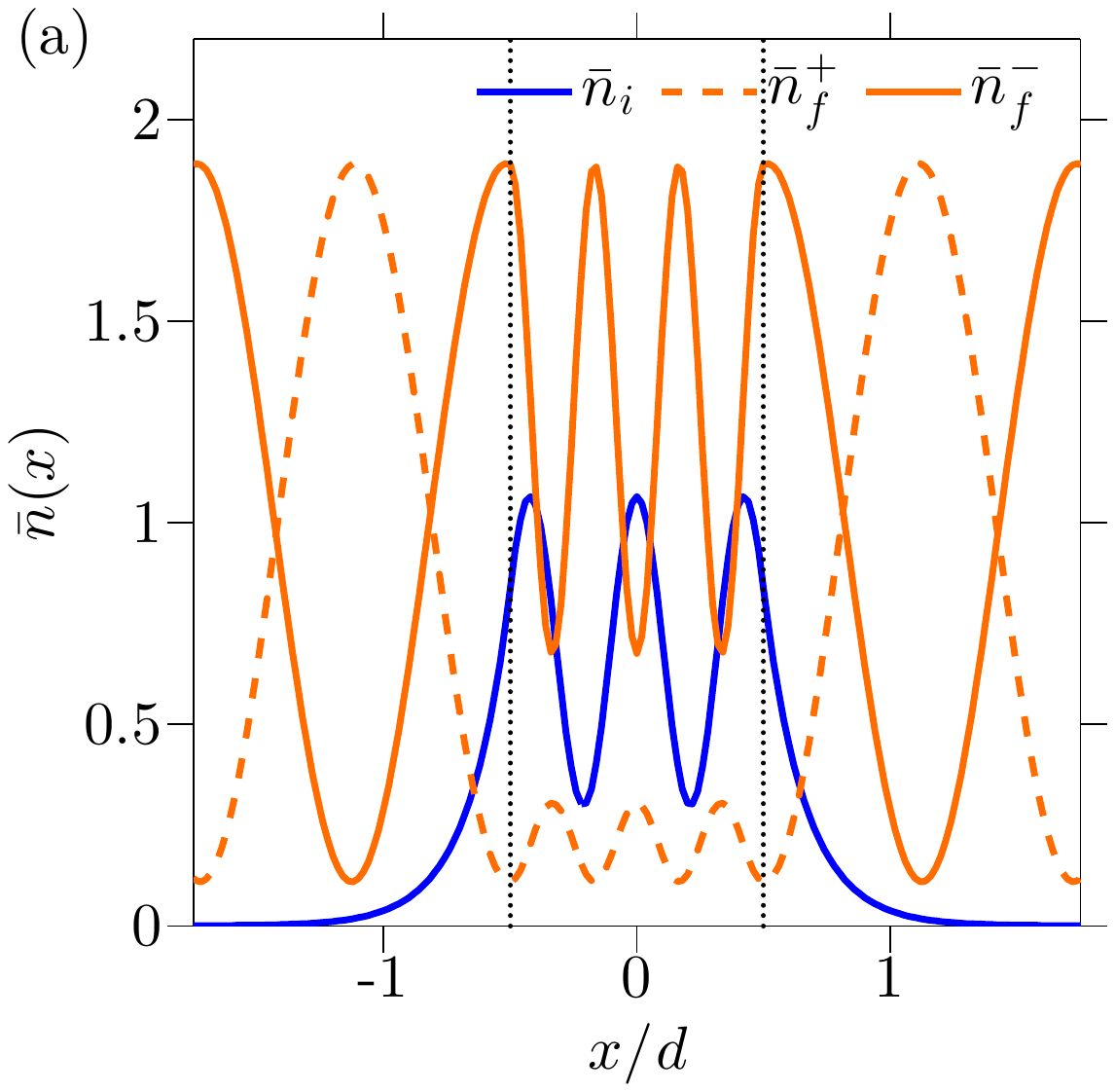}
\includegraphics[width=2.3 in]{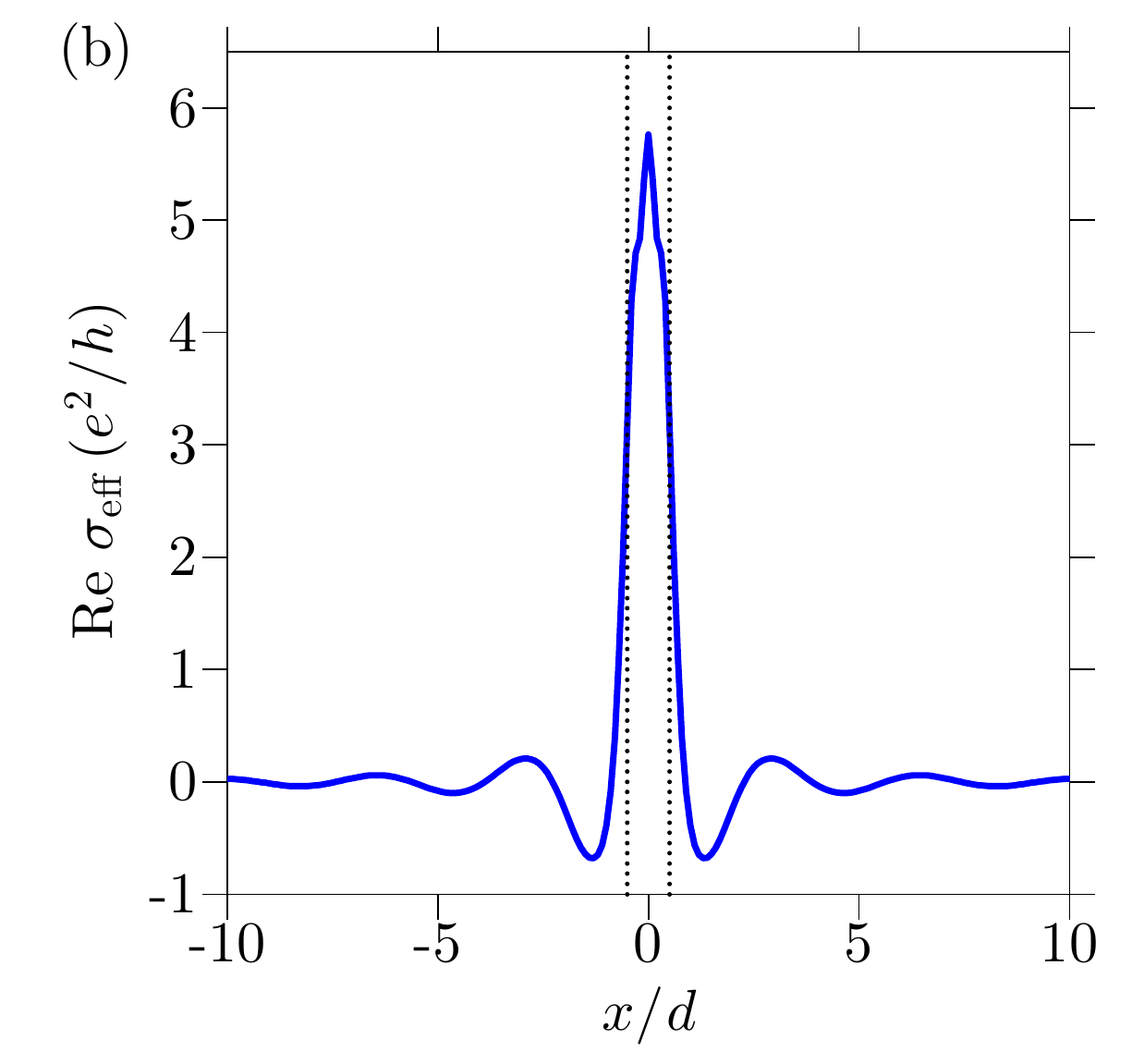}
\includegraphics[width=2.3 in]{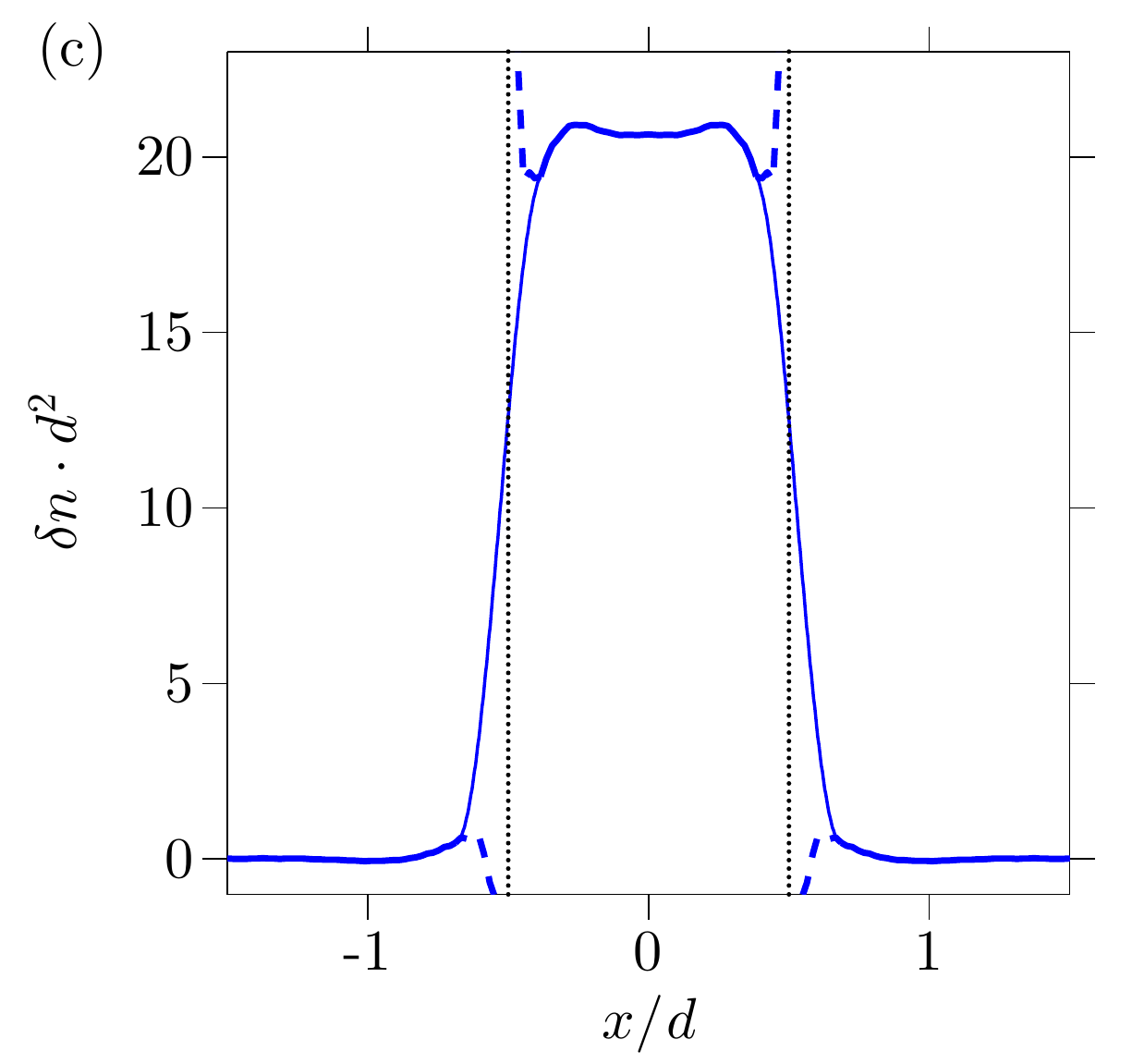}
\caption{
(a) The normalized density distribution $\bar{n}$ of the wave functions in a bound-to-continuum optical transition for $U\equiv ud/\hbar v_F=5$, $K_y\equiv k_y d=5$ and $\omega d/v_F = 1.7$.
The bound state density (blue) is localized at the well, while 
the continuum state densities (orange) are extended.
The bound state is on the $m=3$ branch and has even parity, thus 
transition into the even parity final state (dashed curve) is forbidden.
The vertical dotted lines indicate the boundaries of the square well.
(b) The real part of the effective conductivity calculated with the same parameters as in (a) and $E_F=4$. The peak at the center is caused by bound-to-continuum state transitions, while the long range oscillations with wavevector $\omega/v_F$ come from the transitions between states in the continuum.
(c) Density perturbation induced by a square well of depth $U=5$ at $E_F=4$. The step-like discontinuity in the potential leads to a $1/x$ divergence at the edges of the well, shown by the dashed curves. 
For any realistic smooth potential the divergence is regularized, as shown schematically by the thin solid curves.
}
\label{fig:square_well}
\end{figure*}

We find the complete expression for the wavefunction using the following device.
We imagine that this wavefunction is generated by a plane wave incident from region I, which is then partially transmitted and reflected at each edge of the well. This yields
\begin{equation}
\begin{split}
\Psi^r(\textbf{r}) &= \frac{e^{ik_y y}}{\sqrt{N^c}}
\\
& \times
\begin{cases}
\phantom{\bigg|}\psi_j(\theta)e^{ik_x x}+r_{jh}\,\psi_j\left(\pi-\theta\right)e^{-ik_x x}\,, &\mathrm{I}\\
\phantom{\bigg|}\bar{t}_{jh}\,\psi_h(\bar{\theta})e^{i\bar{k}_x x}+\bar{r}_{jh}\,\psi_h\left(\pi-\bar{\theta}\right)e^{-i\bar{k}_x x}\,, &\mathrm{II}\\
\phantom{\bigg|}t_{jh}\,\psi_j(\theta)e^{ik_x x}\,, &\mathrm{III}
\end{cases}
\end{split}
\label{eqn:square_well_wave_func}
\end{equation}
where $N^c = {L_x L_y}$ is normalization factor equal to the area of the system and $j$ and $h$ can be either $c$ or $v$.
The coefficients $r$, $\bar{t}$, $\bar{r}$ and $t$ are determined by requiring the wavefunction to be continuous at the edges of the well, $x = \pm d / 2$.
If the wavefunction remains in the same band for all three regions, the coefficients are
\begin{align}
& r_{cc}=r_{vv}=i{\dfrac2De^{i(\bar{\varphi}-\varphi)}\sin{\bar{\varphi}}\,(\sin\theta-\sin\bar{\theta})}\,,
\label{eqn:coef_r}\\
&\bar{t}_{cc}=\bar{t}_{vv}={\dfrac2D e^{i\frac{\bar{\varphi}-\varphi}2}\cos\theta\cos\frac{\theta+\bar{\theta}}2}\,,
\label{eqn:coef_t_}\\
&\bar{r}_{cc}=\bar{r}_{vv}={\dfrac2D e^{i\frac{3\bar{\varphi}-\varphi}2}\cos\theta\sin\frac{\theta-\bar{\theta}}2}\,,
\label{eqn:coef_r_}\\
& t_{cc}=t_{vv}={\dfrac2D e^{i(\bar{\varphi}-\varphi)}\cos\theta\cos\bar{\theta}}\,,
\label{eqn:coef_t}
\end{align}
where $\varphi\equiv k_x d$ and $\bar{\varphi}\equiv \bar{k}_x d$, and
\begin{equation} D(k,\,k_y)={1+\cos(\theta+\bar{\theta})-e^{i2\bar{\varphi}}[1-\cos(\theta-\bar{\theta})]}\,.
\end{equation}
If the wavefunction switches band upon entering the well, the coefficients are
\begin{align}
& r_{vc}=r_{cv}=-r_{cc}(\theta\rightarrow\theta-\pi)\,,
\label{eqn:coef_r'}\\
&\bar{t}_{vc}=-\bar{t}_{cv}=\bar{t}_{cc}(\theta\rightarrow\theta-\pi)\,,
\label{eqn:coef_t_'}\\
&\bar{r}_{vc}=-\bar{r}_{cv}=\bar{r}_{cc}(\theta\rightarrow\theta-\pi)\,,
\label{eqn:coef_r_'}\\
& t_{vc}=t_{cv}=t_{cc}(\theta\rightarrow\theta-\pi)\,.
\label{eqn:coef_t'}
\end{align}
If $k_x$ is real,
there is another wavefunction $\Psi^l$ with the same
magnitude $k$ of the momentum (same energy),
which corresponds to the wave incident from region III. 
A quick method to obtain $\Psi^l$ is by reflecting $\Psi^r$ with respect to the $y$-axis:
\begin{equation}
\Psi^l(x) = \sigma_y \Psi^r(-x)\,.
\end{equation}
From $\Psi^r$ and $\Psi^l$ we construct the orthogonal eigenstates
\begin{equation}
\Psi^\pm = \dfrac{\Psi^r\pm\Psi^l}{\sqrt{2}}\,,
\end{equation}
which are labeled by their parity $P$:
\begin{equation}
\Psi^P(x) = P \sigma_y \Psi^P(-x)\,,
\quad P = \pm1\,.
\end{equation}
From the above expression we deduce that states localized within the potential well must also exist.
Indeed, whenever $|\bar{k}|>|k|$ there exist states with $|k|<|k_y|$,
so that $k_x$ is imaginary and the wavefunction is evanescent outside the well. 
This happens when the denominators vanish, $D = 0$,
so that the eigenstate exists without an incident plane wave from outside the well.
The dispersion of these bound states is found by solving
\begin{equation}
i(k \bar{k} - k_y^2)\tan\bar{\varphi} = k_x \bar{k}_x\,.
\label{eqn:loc_dispersion2}
\end{equation}
The wavefunction still has the form of Eq.~\eqref{eqn:square_well_wave_func}, but with a normalization factor
\begin{equation}
N^b = 2 L_y d \left[
|\bar{t}|^2 \left(1+P\sin\bar{\theta}\,\dfrac{\sin\bar{\varphi}}{\bar{\varphi}}\right)
+ |t|^2\sin\theta\,\dfrac{e^{-q d}}{2 q d}
\right]\,.
\end{equation}
Note that ${k}_x$ is now imaginary, $k_x \equiv iq$ with $q > 0$. 
Each branch of solution except the one terminating at $k = k_y = 0$ is the continuation of Fabry--P\'{e}rot (FP) modes outside the continuum. 
The FP modes correspond directly to the supercritical or quasi-bound states.
They satisfy the
resonance condition $\bar{\varphi} = \pi m$ with $m=1,2,\ldots$,
so branches of smaller $m$ emerge at higher $k_y$.
%The merging point 
This condition can be expanded to find 
the expression for the critical point at which the $m$th bound state emerges from the valence band,
% is determined by the following relation between the well depth $u$ and $k_y$,
%%
\begin{equation}
	\frac{u}{\hbar v_F} = |k_y| + \sqrt{k_y^2 + (m\pi/d)^2}\,.
	\label{eqn:U_c}
\end{equation} 
All the bound state branches asymptotically approach the line $k = |k_y| - u / \hbar v_F$ as $|k_y| \to \infty$.
The wavefunction of each branch is alternatively even ($P=+1$) or odd ($P=-1$) with the lowest branch being even. 
An example of the normalized density  of a bound state $\bar{n}\equiv|\Psi|^2 L_y d$ is shown in Fig.~\ref{fig:square_well}(a) along with the normalized density of continuum states $\bar{n}\equiv|\Psi|^2L_x L_y$ for comparison.

%For a given $K_y$, 

Having found the expression for the eigenstate wavefunctions, we use the Kubo formula to calculate the nonlocal conductivity,
\begin{equation}
\begin{split}
\sigma(\textbf{r},\,\textbf{r\ensuremath{^{\prime}}})=
-\dfrac{1}{i\omega}&\sum_{i,f}\dfrac{\nu_f-\nu_i}{\varepsilon_f-\varepsilon_i-(\omega+i0^+)}\times
\\
&\big\langle\Psi_i(\textbf{r})|\,\hat{j}_x\,|\Psi_f(\textbf{r})\big\rangle\big\langle\Psi_f(\textbf{r\ensuremath{^{\prime}}})|\,\hat{j}_x\,|\Psi_i(\textbf{r\ensuremath{^{\prime}}})\big\rangle\,,
\end{split}
\label{eqn:kubo_formula}
\end{equation} 
where $i$ and $f$ represent initial and final states, $\nu$ is the Fermi-Dirac occupation factor of the state with $\nu_i=1$ and $\nu_f=0$, $\varepsilon=\hbar v_F k$ is the energy of the state, and $\hat{j}_x=ev_F{\sigma_z}$ is the current operator.
%The initial states can be localized states or continuum states, while the final states are continuum states.
%Transition from localized to localized states has no contribution due to the symmetry of the states. 
Assuming that the total field $\textbf{E}$ is uniform and parallel to $\hat{x}$, $\textbf{E}= E_x \hat{x},$ the current-field relation can be simplified to
\begin{equation}
j_x(x) =
\int\limits_{-\infty}^{\infty}d\textbf{r\ensuremath{^{\prime}}}\,E_x
\sigma(\textbf{r},\,\textbf{r\ensuremath{^{\prime}}})
\equiv E_x \sigma_\mathrm{eff}(x)\,.
\label{eqn:sigma_eff_def}
\end{equation}
Note that the integral over $y'$ enforces the conservation of $k_y$,
\begin{equation}
\int\limits_{-\infty}^{\infty}dy'\,e^{ik_{yi}}e^{-ik_{yf}}
=2\pi\delta(k_{yi}-k_{yf})\,,
\end{equation}
so that all $y$ and $y'$ dependences cancel out in Eq.~\eqref{eqn:sigma_eff_def}.
%while the current operator becomes $\hat{j}=\sigma_z$.
We are interested in the real part of the effective conductivity.
Transitions that contribute satisfy the relations 
\begin{equation}
k_{yf}=k_{yi}\,,\quad k_f=k_i+\dfrac{\omega}{v_F}\,.
\label{eqn:transition_condition}
\end{equation} 
Additionally, the $i/f$ states must have the opposite parity as 
the matrix element
\begin{equation}
M(x)\equiv\big\langle\Psi_i(x)|\,\sigma_z\,|\Psi_f(x)\big\rangle
\end{equation}
is odd in $x$ when the $i/f$ states have the same parity.
%Hence there is nonzero contribution to the real part of the effective conductivity only when the $i$ and $f$ states have opposite symmetry.
To proceed, we impose periodic boundary conditions and extend the system size to infinity, so that
\begin{equation}
\sum_{i,f}\rightarrow g \times
\begin{cases}
\dfrac{L_y}{2\pi} {\displaystyle\int} dk_y\,,
&(\mathrm{bound})
\\[12pt]
\dfrac{L_x L_y}{(2\pi)^2} {\displaystyle\int} d k_x d k_y\,,
&(\mathrm{continuum})
\end{cases}
\end{equation}
where $k_y$ is taken to be positive and $g=8$ is the total degeneracy, including spin, valley and contribution from negative $k_y$.
Applying the Sokhotski--Plemelj formula
\begin{equation}
\mathrm{Im}\,\dfrac{1}{x - i 0^+} = \pi\delta(x)
\end{equation}
to Eqs.~\eqref{eqn:kubo_formula} and \eqref{eqn:sigma_eff_def}, 
we find for the bound-to-bound state transitions,
\begin{equation}
\begin{split}
\mathrm{Re}\,\sigma_\mathrm{eff}^{bb}(x) &= g \pi \frac{e^2}{h}\frac{v_F}{\omega} {L_y^2}
\\
&\times
\left|\dfrac{dk_i}{dk_y}-\dfrac{dk_f}{dk_y}\right|^{-1}_{k_y^*}
 M(x) \int\limits_{-\infty}^{\infty} dx'\,M^{*}(x')\,,
\end{split}
\end{equation}
where $k_y^*$ satisfies Eq.~\eqref{eqn:transition_condition}.
For bound-to-continuum transitions, we get
\begin{equation}
\begin{split}
\mathrm{Re}\,\sigma_\mathrm{eff}^{bc}(x) &= \frac{g}2\frac{e^2}{h} \frac{v_F}{\omega} L_x L_y^2
\\
&\times
\int\limits_{k_y^\mathrm{min}}^{k_y^\mathrm{max}}
 \frac{dk_{yi}}{|\cos\theta_f|} M(x)
\int\limits_{-\infty}^{\infty}dx'\,
M^*(x')\,.
\end{split}
\end{equation}
The limits of $k_y$ are determined from the dispersion, the frequency $\omega$ and the doping level $k_F = \mu / (\hbar v_F)$. 
%Note that the final state have the following relation to the initial state,
Continuum-to-bound state transitions result in the same expression except that the labels $i$ and $f$ are interchanged.
The resultant conductivity has a peak around the well and decays quickly away from the well [Fig.~\ref{fig:square_well}(b)].
Finally,
continuum-to-continuum transitions yield
\begin{equation}
\begin{split}
\mathrm{Re}\, \sigma_\mathrm{eff}^{cc}(x) &= \frac{g}{4\pi}\frac{e^2}{h}\frac{v_F}{\omega}L_x^2 L_y^2
\\
&\times 
\int\limits_{0}^{k_F}dk_{xi}\int\limits_{k_y^\mathrm{min}}^{k_y^\mathrm{max}}
 \frac{dk_{yi}}{|\cos\theta_f|}M(x)
\int\limits_{-\infty}^{\infty}dx'\,
M^*(x')\,,
\end{split}
\end{equation}
where 
\begin{equation}
\begin{split}
& k_y^\mathrm{min}=\sqrt{\mathrm{max}\left[0,\,\left(k_F-\frac{\omega}{v_F}\right)^2-k_{xi}^2\right]}\,,
\\
& k_y^\mathrm{max}=\sqrt{k_F^2-k_{xi}^2}\,.
\end{split}
\end{equation}
This results in an oscillating conductivity with a period of $2\pi v_F/\omega$ which can be negative, that is, the local current in real space can go in the opposite direction as the field.
This is however no cause for alarm.
Consider the case of uniform undoped graphene where the conductivity
as a function of momentum at fixed $\omega$ is
\begin{equation}
\sigma_0(q) = -i\dfrac{e^2}{4\hbar}
\dfrac{\omega}{\sqrt{v_F^2 q^2 - \omega^2}}\,.
\end{equation}
The corresponding conductivity in the real-space is
\begin{equation}
\sigma_0(x) = \int\limits_{-\infty}^{\infty} \frac{d q_x}{2\pi}\,
\sigma_0(q_x) e^{i q_x x}
= \frac{e^2}{8\hbar} \frac{\omega}{v_F}
H_0^{(1)}\left(\frac{\omega}{v_F}x\right)\,,
\end{equation}
so that $\mathrm{Re}\,\sigma_0(x) \propto J_0\bigl(\frac{\omega}{v_F} x\bigr)$ can be negative.
Thus conductivity oscillations with period $\sim 2\pi v_F/\omega$ is a general property in the presence of nonuniformity.
For $\omega = 890\unit{cm}^{-1}$ in our s-SNOM experiment
this period is $37\unit{nm}$ and such oscillations cannot be easily resolved. 
Therefore, we draw the reader's attention to another
feature of the computed local conductivity, which is the prominent 
peak near the origin, see
Fig.~\ref{fig:square_well}(b).
We conclude that our simple model does predict a strong enhancement of dissipation near the nanotube, in agreement
with the experiment.
The extra dissipation is caused by the bound-to-bound state optical transitions.
Of course, these calculations are not meant to be quantitatively compared
with the experiment because our model of the square-well potential is not fully realistic.
The quantity more suitable for the purposes of a qualitative comparison is
the average value of the effective conductivity
\begin{equation}
\bar{\sigma} \equiv \frac{1}{W}\, \int\limits_{-\infty}^{\infty} d x\,\mathrm{Re}\,\sigma_\mathrm{eff}(x)\,.
\label{eqn:bar_sigma}
\end{equation}
similar to Eq.~(4) of the main text.
As long as $W$ is larger than $d$ but smaller than the spatial resolution,
the near-field profile is sensitive only to the product $\bar{\sigma} W$
not the precise value of $W$, see Appendix~\ref{sec:fitting}.
The results of our calculations of $\bar{\sigma}$ are shown in Fig.~3 of the main text.
For simplicity, in these calculations we excluded the part of $\mathrm{Re}\,\sigma_\mathrm{eff}$ resulting from continuum-to-continuum transitions
because it is relatively small at the potential well.

%%%%%%%%%%%%%%%%%%%%%%%%%%%%%%%%%%%%%%%%%%%%%%%%%%%%%%%%%%%%%%%%%%%%%%%%%%%
\section{Local density and density of states}
\label{sec:LDOS}

\begin{figure*}
\includegraphics[height=2.3 in]{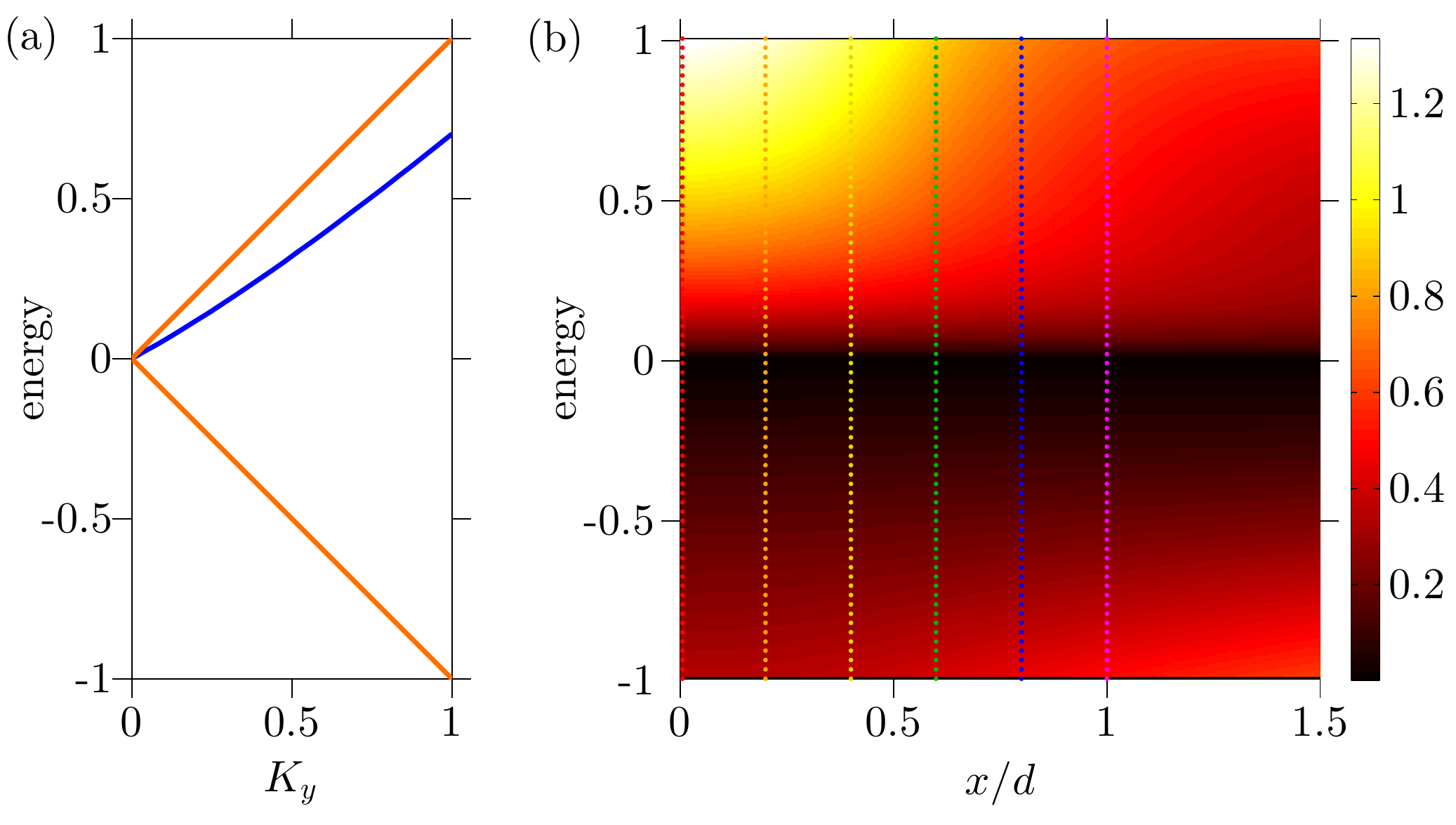}
\includegraphics[height=2.3 in]{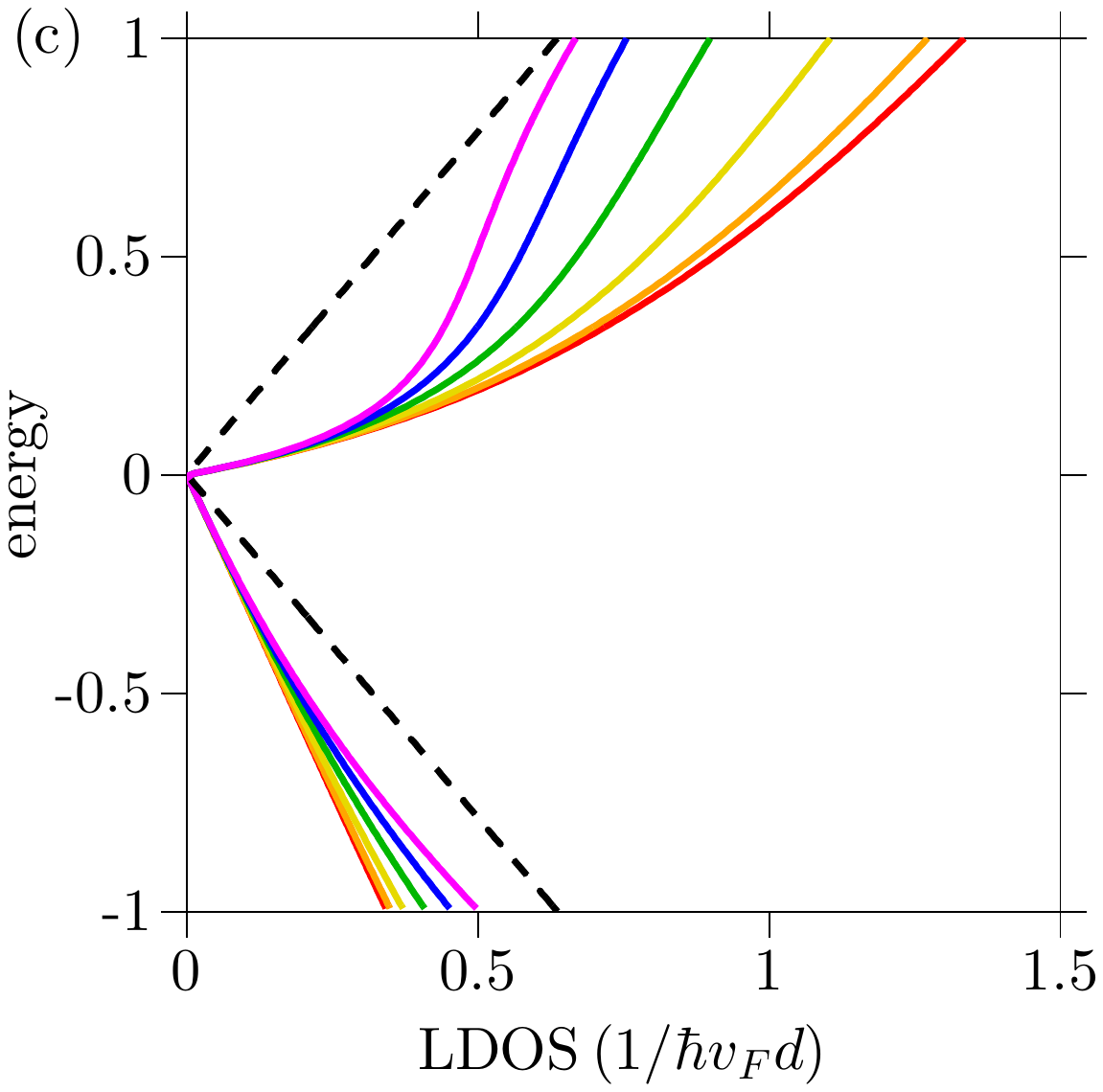}
\\
\includegraphics[height=2.3 in]{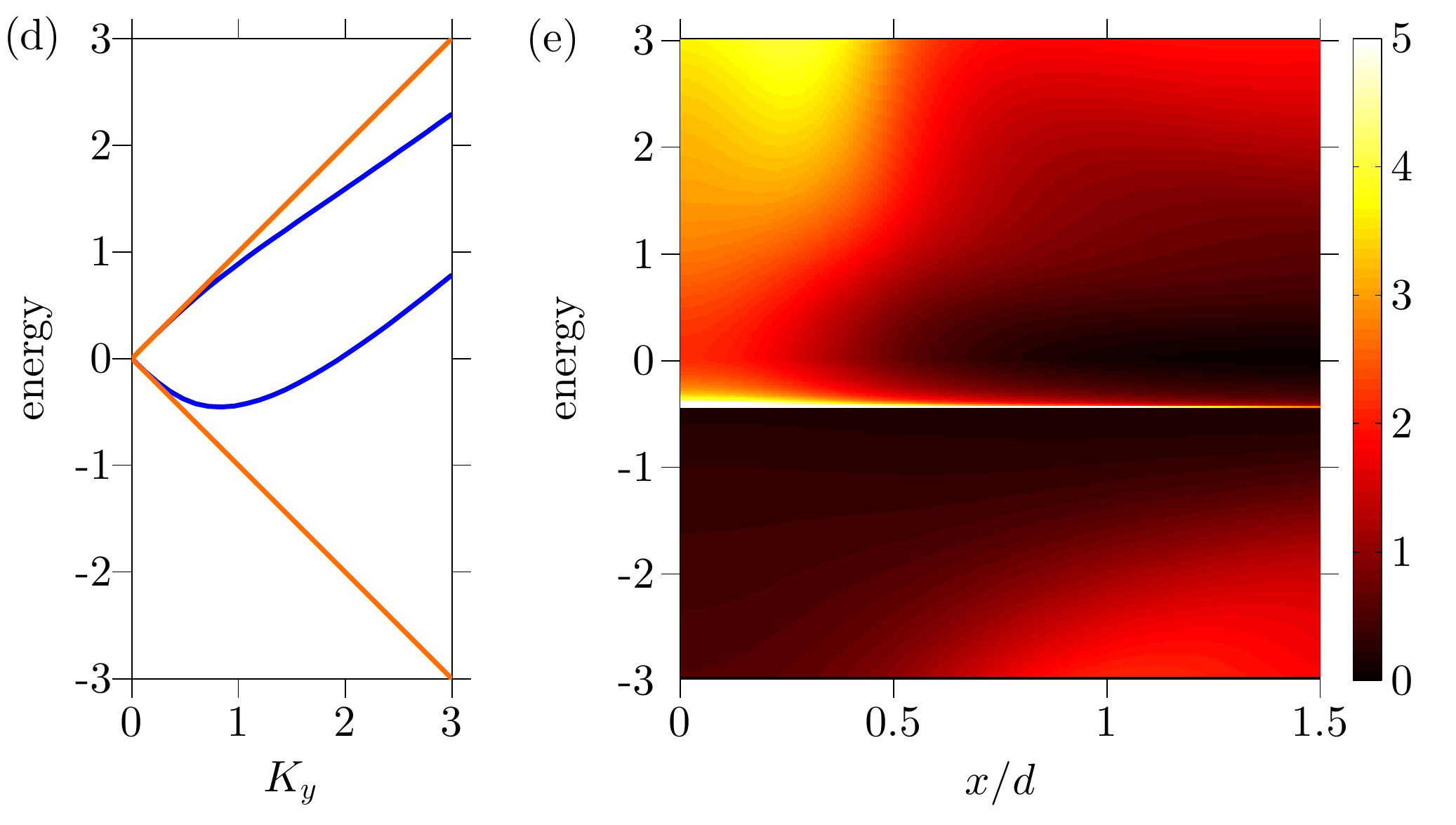}
\includegraphics[height=2.3 in]{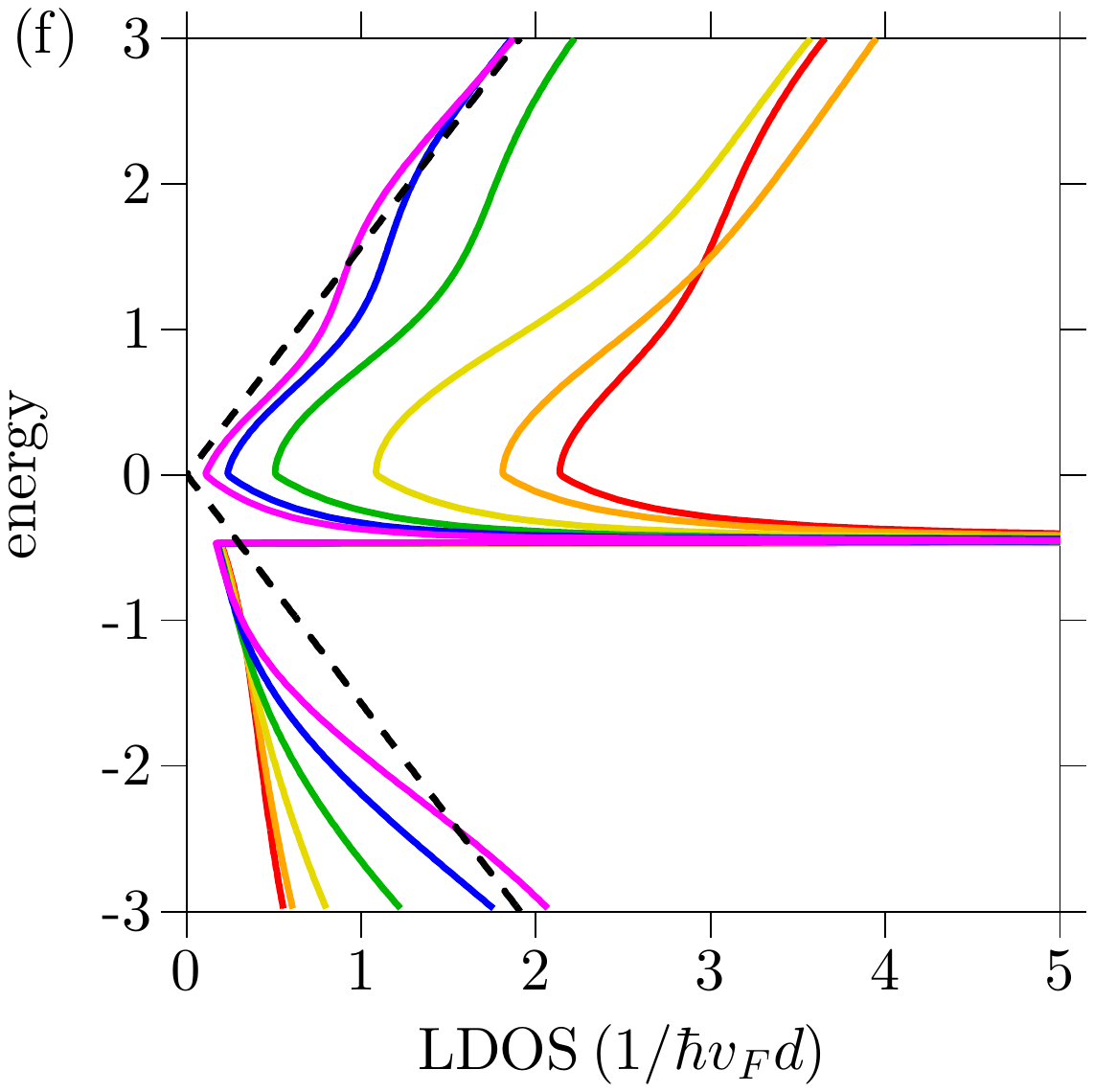}
\\
\includegraphics[height=2.3 in]{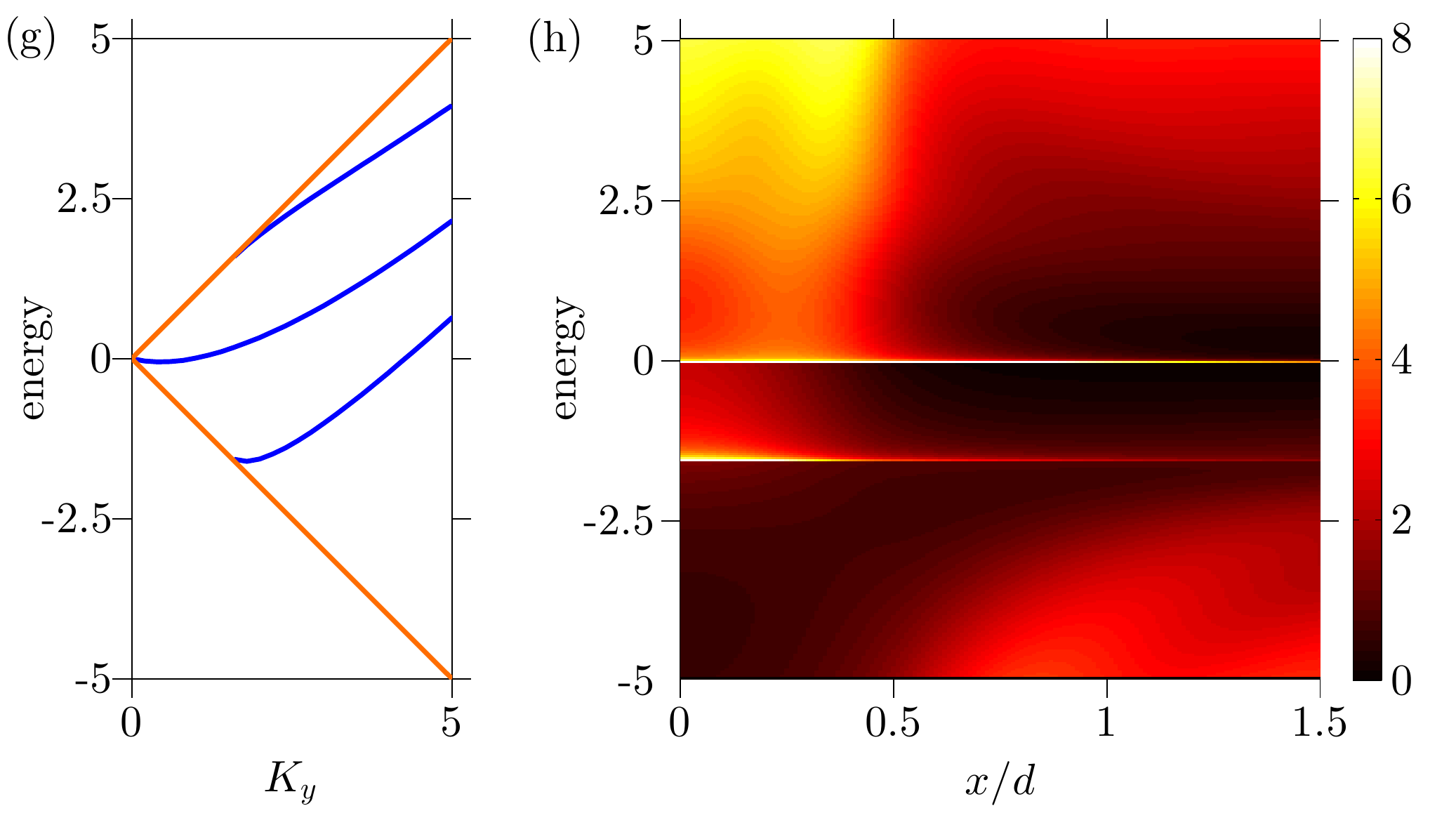}
\includegraphics[height=2.3 in]{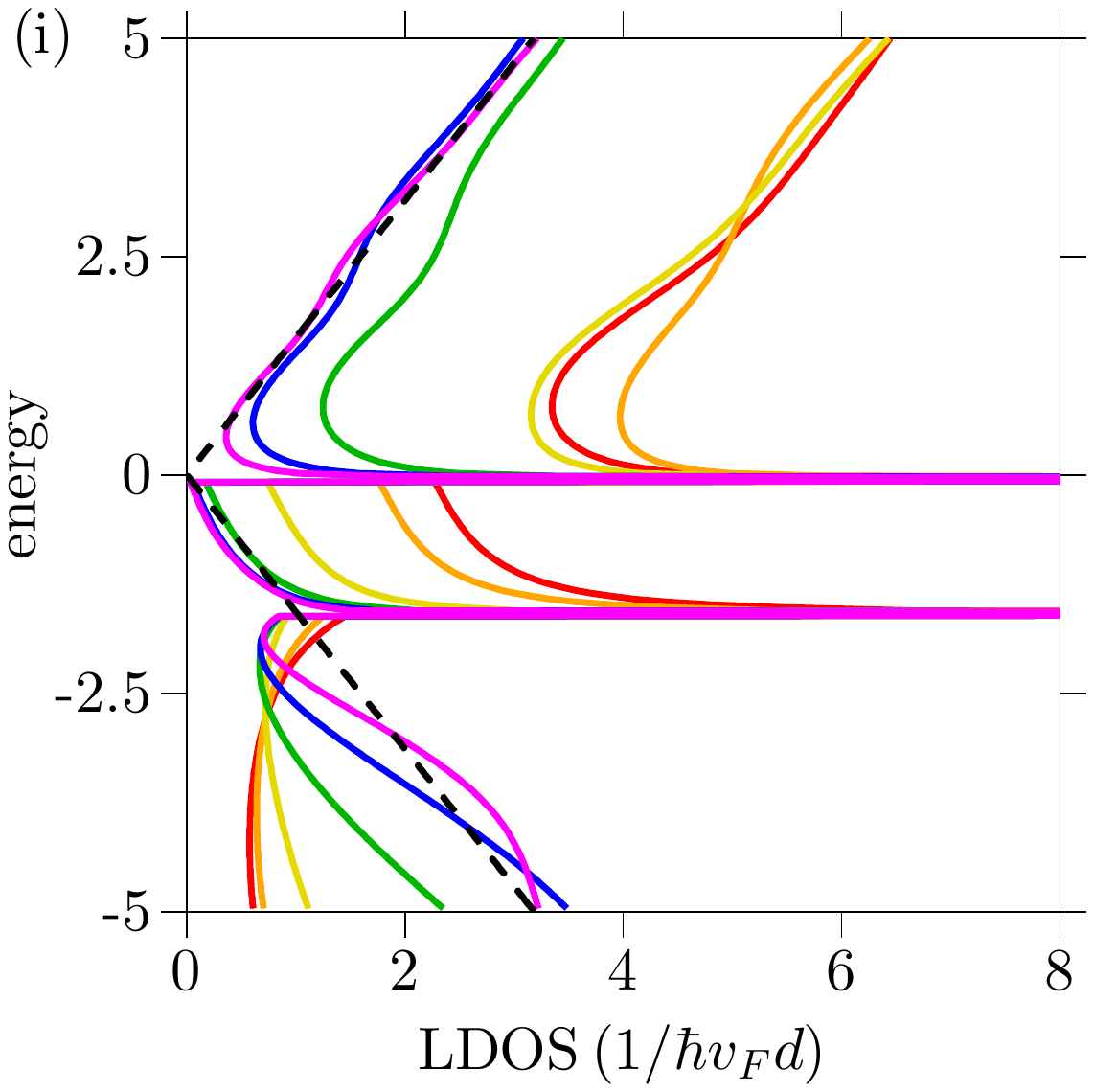}
\caption{
The local density of states $\nu$ in the vicinity of a square well potential.
(a) The dispersion of the bound states for well depth $U=1$.
(b) A false color plot of the LDOS as a function of the dimensionless energy $E = k d$ and distance $x / d$.
(c) Cross sections of (b) taken at several distances from $x=0$ to $x=d$.
The dashed black lines represent the LDOS of unperturbed graphene, $\nu_0=2|k|/\pi\hbar v_F$.
Similar quantities are shown in (d)-(f) for $U=3$ and in (g)-(i) for $U=5$.
}
\label{fig:LDOS_supp}
\end{figure*}

For better understanding of the effect of the potential well on electronic properties it is instructive to consider two other local observables: the carrier density and the local density of states.
We begin with 
the density perturbation $\delta n(x)$.
To find this quantity we
first find the square of the absolute value of the wavefunctions $|\Psi(x, y)|^2$ of the occupied eigenstates under the potential well, integrated over $k_y$ and $k$.
The integration
is done over the energies $k$ bounded from above by the Fermi energy and from below by a cutoff energy $k_m$, a large negative number.
The same procedure is then repeated for the unperturbed eigenstates (without the potential well).
The difference of the two results is $\delta n$.
There is however a complication to this procedure rooted in the ``chiral anomaly'' in the quantum field-theory of free Dirac fermions.
For $x$ inside the well, the lower cutoff $k_m$ for the unperturbed eigenstates must be changed to $\bar{k}_m = k_m - u /\hbar v_F$.
Without this redefinition, $\delta n(x)$ would diverge as
$k_m$ is decreased.
Once this proper background subtraction is done, the integration converges to a finite value $\delta n(x)$ everywhere except at the edges of the well, $x \to \pm d / 2$.
The remaining divergence can be traced to the discontinuity in the potential $v(x)$.
To arrive at this conclusion we reasoned that the divergence is produced by large negative energies, and so it could be investigated using the perturbation theory.
Therefore, we considered the linear-response theory expression for
the density perturbation:
\begin{equation}
\delta n(x) = \int \frac{d q}{2\pi} e^{i q x} \Pi(q) v(q)\,,
\end{equation}
where
\begin{equation}
v(q) = -\frac{2 U_0} q \sin\left(\frac{q d}{2}\right)
\end{equation}
is the Fourier transform of the potential $v(x)$ and
$\Pi(q)$ is the static polarization function of graphene.
At large $q$ this function behaves as
\begin{equation}
\Pi(q) = -\frac{|q|}{4\hbar v_F}
\end{equation}
regardless of the doping level.~\cite{Wunsch2006dpo}
Evaluation of the integral for $\delta n$ using this asymptotic form yields
\begin{equation}
\frac{u}{4\pi\hbar v_F}
\left(\frac1{x + \frac{d}{2}}-\frac1{x - \frac{d}{2}}\right)\,,
\end{equation}
which matches the numerical results calculated as described above for undoped graphene [Fig.~\ref{fig:square_well}(c)].
In reality, i) the linear dispersion of Dirac fermions does not extend to infinite momenta and ii) the potential must be smooth.
Either way
the divergence is regularized and the perturbed density is smooth as well,
see Fig.~\ref{fig:square_well}(c).
Not surprisingly, this box-like density profile is different from the more realistic
Lorentzian function [Eq.~\eqref{eqn:n}] we used to fit the experimental data in the main text and in Appendix~\ref{sec:fitting} below.
However, as we argued in Appendix~\ref{sec:square_well},
a qualitative comparison between the present model and the experiment is still meaningful.

Let us now turn to the local density of states (LDOS) $\nu$.
Previously, the LDOS of graphene around clusters of pointlike charged impurities has been measured by scanning tunneling spectroscopy.~\cite{Wang2013oac}
These experiments discovered peaks in LDOS,
which were attributed to the emergence of the supercritical quasi-bound states.~\cite{Shytov2007acq}
We find that for a 1D perturbation the bound states, instead of the quasi-bound ones, produce the dominant features in the LDOS.

The contribution of the bound states to the LDOS is given by
\begin{equation}
\nu(k, x) = \frac{g}{2\pi}
\frac{L_y}{\hbar v_F}
\sum_{i} \left|\frac{d\tilde{k}_{yi}}{dk}\right|
\left|\Psi(\tilde{k}_{yi},x)\right|^2\,,
\end{equation}
where $\tilde{k}_{yi}$ are positive solutions of Eq.~\eqref{eqn:loc_dispersion2} at a given $k$.
The contribution of the delocalized states in the continuum is
\begin{equation}
\nu(k,x) = \frac{g}{4\pi^2}
\frac{L_x L_y}{\hbar v_F}\, |k|
\sum_{P = \pm} \int_{0}^{\pi/2} d\theta \left|\Psi^P(k,\theta,x)\right|^2\,.
\end{equation}
We show in Fig.~\ref{fig:LDOS_supp}(a) the dispersion of bound states and in Fig.~\ref{fig:LDOS_supp}(b) the false color plot of the LDOS for the case of $U=1$.
The bound states contribute to the significant increase in $\nu$ at positive energies and for distances $x$ inside the well.
This is seen more clearly in Fig.~\ref{fig:LDOS_supp}(c), where vertical cross sections of the false color plot is taken at several distances inside and outside the well. 
The contribution of the bound states drops quickly outside the well and approaches the unperturbed LDOS $\nu_0 = 2|k| / (\pi\hbar v_F)$ shown in dashed lines.
Similar plots for cases $U=3$ and $U=5$ are shown in Fig.~\ref{fig:LDOS_supp}(d)-(f) and in (g)-(i), respectively.
In these two cases the LDOS inside the well are similarly increased due to the bound states.
However, the most prominent features of the LDOS are the van Hove singularities that
are caused by the extrema in the bound state dispersions.
For $U=3$ the singularity occurs at dimensionless energy $E = k d = 0.45$, while for $U = 5$ they occur at $E = -0.05$ and $E = -1.59$.
A quasi-bound state is present for the case of $U = 5$, whose contribution is manifest as the increase of the LDOS inside the well just before the van Hove singularity at $E = -1.59$, as shown in Fig.~\ref{fig:LDOS_supp}(i).
The former is a relatively weak feature in comparison to the latter.

We think that these properties of the LDOS should be quite generic for hypercritical potentials in graphene.
Therefore, despite the oversimplification of the square-well potential model,
our analysis may provide a useful reference for future scanning tunneling experiments with such potentials.

%%%%%%%%%%%%%%%%%%%%%%%%%%%%%%%%%%%%%%%%%%%%%%%%%%%%%%%%%%%%%%%%%%%%%%%%
%%
\begin{figure}[b]
	\includegraphics[height=1.75in]{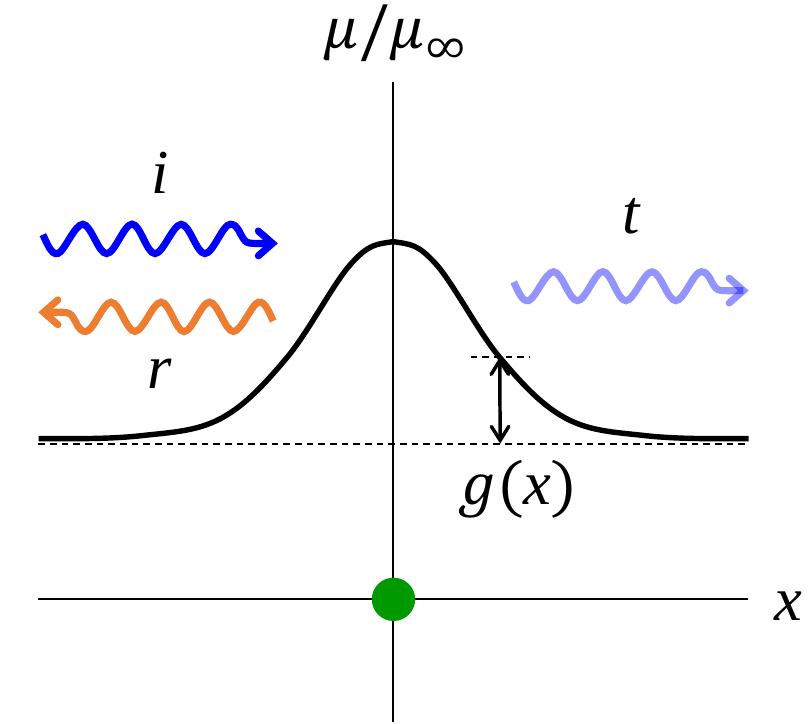}
	\caption{Schematic of an incident, reflected, and transmitted plasmon wave near
		the electronic inhomogeneity $g(x)$ caused by the linelike charge perturbation (green).
	}
	\label{fig:plasmon_refl}
\end{figure}

\section{Plasmon reflection from a linelike charge perturbation}
\label{sec:r}

In this section we summarize the theory of
plasmon reflections from a linelike charge perturbation.~\cite{Fei2013epp}
At this stage we are not yet discussing how the incident plasmon wave is created or how the reflected wave can be measured.
Those questions are addressed in Appendix~\ref{sec:fitting}
devoted to realistic simulations of s-SNOM experiment.
Here our purpose is to specify the model assumptions and to present
the analytical results.

Our main assumption is that we can
describe the response of graphene by a local conductivity $\sigma(x)$.
This assumption is readily justified if
the density $n(x)$ and the chemical potential $\mu(x)$ of graphene are smoothly varying,
see Fig.~\ref{fig:plasmon_refl}.
Thus, if the plasmon energy $\hbar\omega$ is much smaller than $\mu$ everywhere,
$\sigma(x)$ is given by [Eq.~(5) of the main text],
\begin{equation}
	\sigma(x) = \frac{i}{\pi\omega}\,
	\frac{\mathcal{D}(x)}{1 + i \gamma(x)}\,,
	\quad
	\mathcal{D}(x) = \frac{e^2}{\hbar^2}\, |\mu(x)|\,.
	\label{eqn:graphene_conductivity}
\end{equation}
Here $\mathcal{D}$ is the Drude weight and the dimensionless function
$\gamma$ is the phenomenological damping rate.
We assume that the system remains uniform in the $y$-direction at all $x$.
The legitimacy of the local conductivity approximation is less obvious if
the carrier density varies sharply, e.g.,
in a box-like fashion depicted in Fig.~\ref{fig:square_well}(c).
However, it should be indeed valid in the context of the plasmon propagation
if the plasmon wavelength is longer than the characteristic length scale of nonlocality (the Fermi wavelength or the characteristic width of the inhomogeneity, whichever is larger).
The
effective local conductivity can then be defined by averaging the nonlocal one over a suitable interval $W$, see Eq.~\eqref{eqn:bar_sigma}.
In this case, Eq.~\eqref{eqn:graphene_conductivity} should be considered a formal parametrization of function $\sigma(x)$.
In particular, $\gamma(x)$ should be understood as
damping averaged over the lengthscale $W$.

Let us suppose that at $x\to \pm \infty$, $n(x)$ and $\mu(x)$ approach constant values $n_\infty$ and $\mu_\infty$, respectively and
let us define the dimensionless function
\begin{equation}
	g(x) = \frac{\sigma(x)}{\sigma_\infty} - 1\,,
	\quad
	\sigma_\infty \equiv \sigma(\infty)\,.
	\label{eqn:g_def}
\end{equation}
If $\gamma$ were constant, this function would be equal to $g = |\mu / \mu_\infty| - 1$,
see Fig.~\ref{fig:plasmon_refl}.

We want to study how an incident plasmon plane wave with momentum
$(q_x, q_y)$ is scattered by the inhomogeneity. We will show that the corresponding reflection coefficient is given by the formula
\begin{equation}
	r \simeq i \dfrac{q_x^2 - q_y^2}{q_x} \tilde{g}(-2q_x)\,,
	\label{eqn:r_FBA}
\end{equation} 
where
\begin{equation}
	\tilde{g}(k) = \int\limits_{-\infty}^{\infty} d x\, g(x)e^{-i k x}\,.
\end{equation} 
In particular, for normal incidence, $q_x = q_\infty$, $q_y = 0$,
the reflection coefficient is
\begin{equation}
	r \simeq i q_\infty \tilde{g}(-2 q_\infty)\,,
	\label{eqn:r_normal}
\end{equation} 
similar to the usual first Born approximation.
Note that because of the translational invariance in $y$, the momentum $q_y$ is conserved.

Let us outline the derivation. Assuming $q \gg \omega / c$, which is satisfied in our experiment, we can neglect retardation and treat the problem in the quasistatic approximation.
As our main dependent variable we choose the electric potential
$\Phi = \Phi(\mathbf{r})$.
In general, $\Phi$ is the sum of the external potential $\Phi_\mathrm{ext}$ and
the potential induced by charge density $\rho$ in graphene,
\begin{equation}
	\Phi(\textbf{r}) = \Phi_\mathrm{ext}(\textbf{r})
	+ (V \ast \rho)(\textbf{r})\,,
	\label{eqn:coulomb_law}
\end{equation}
where $V(\textbf{r}) = 1/\kappa r$ is the Coulomb kernel and the asterisk denotes convolution,
\begin{equation}
	(A \ast B)(\textbf{r}) \equiv \int d^2 r'\, A(\textbf{r}-\textbf{r}')B(\textbf{r}')\,.
\end{equation}
Combining together these equations plus the continuity equation for current and charge density,
we obtain
\begin{equation}
	\Phi(\textbf{r}) = \Phi_\mathrm{ext}(\textbf{r})
	- V(\textbf{r})\ast{\nabla}
	\left(\frac{\sigma(\textbf{r})}{i\omega}{\nabla}\Phi(\textbf{r})\right)\, .
	\label{eqn:plasmon_eq}
\end{equation}
For an ideal uniform sample the solution of this equation has the form of a Fourier integral:
\begin{equation}
	\Phi(\textbf{r}) = \int \frac{d^2 q}{(2\pi)^2}\, e^{i \textbf{q} \textbf{r}}
	\dfrac{\tilde{\Phi}_\mathrm{ext}(\textbf{q})}{\epsilon(q)}\,,
	\quad
	\epsilon(q) = 1 -\dfrac{q}{q_p} 
	\,.
	\label{eqn:Phi}
\end{equation} 
The zero of the dielectric function $\epsilon(q)$ defines the plasmon momentum
\begin{equation}
	q_p = \frac{i \kappa \omega}{2 \pi \sigma}
	\label{eqn:q_p}
\end{equation} 
introduced in the main text.
The momentum $q_p$ is complex
for any finite damping, $\gamma > 0$, with $\mathrm{Im}\, q_p > 0$ having the physical meaning of the inverse propagation length.
Indeed, in the absence of the external potential,
one can find (unbounded) solutions $\Phi = e^{i q_x x + i q_y y}$
with real $q_y$ and complex $q_x = \sqrt{q_p^2 - q_y^2}$, $\mathrm{Im}\, q_x > 0$,
which can be thought of as decaying plane waves that are incident from the far left at some oblique angle.
In the problem we study graphene is inhomogeneous, $q_p$ is $x$-dependent,
\begin{equation}
	\frac{1}{q_p(x)} = \frac{1+g(x)}{q_\infty}\,,
	\quad q_\infty \equiv q_p(\infty)\,,
	\label{eqn:g_def2}
\end{equation}
and so the solution would contain 
the incident and the scattered (reflected plus transmitted) waves,
see Fig.~\ref{fig:plasmon_refl}.

Setting ${\Phi}_\mathrm{ext}(\mathbf{r}) \to 0$ and
$\Phi(\mathbf{r}) \to \Phi(x) e^{i q_y y}$ in Eq.~\eqref{eqn:plasmon_eq},
we obtain the equation for $\Phi(x)$:
\begin{equation}
	\Phi(x) = V_1 \ast \left(\dfrac{1 + g(x)}{q_\infty} q_y^2 \Phi(x)
	- \partial_x \dfrac{1+g(x)}{q_\infty} \partial_x\Phi(x)
	\right)\,.
	\label{eqn:plasmon_eq_1d}
\end{equation}
Here $V_1(x) = K_0(|q_y x|) / \pi$ is the 1D Coulomb kernel and $K_0(z)$ is the modified Bessel function of the second kind.
Using the Green's function
\begin{equation}
	G(x, q_y) = \int_{-\infty}^{\infty} \frac{dk}{2\pi} e^{i k x}
	\epsilon^{-1}\left(\sqrt{k^2 + q_y^2}\right)\,,
\end{equation} 
we find the equation for the scattered wave
$\psi \equiv \Phi(x) - e^{i q_x x}$:
\begin{equation}
	\psi(x) = q^{-1}_\infty (G\ast V_1) \ast \left(g(x) q_y^2 \Phi(x)
	- \partial_x g(x) \partial_x \Phi(x)\right).
	\label{eqn:plasmon_eq_1d_G}
\end{equation}
We expect $\psi(x) \simeq r e^{-i q_x x}$
at large negative $x$, which implies
\begin{equation}
	r = -\frac{i}{q_x} \int\limits_{-\infty}^\infty d x\, e^{i q_x x} \left\{
	q_y^2 g(x) \Phi(x) - \partial_x \left[g(x) \partial_x \Phi(x) \right]\right\}.
\end{equation} 
To the first order in the small parameter $g(x)$ we can replace $\Phi(x)$ with $e^{i q_x x}$ in the integral, which leads to Eqs.~\eqref{eqn:r_FBA}.
A particularly simple result is obtained if the plasmon wavelength
\begin{equation}
	\lambda_\infty = 2\pi / q_\infty
	\label{eqn:lambda}
\end{equation} 
is much larger than the 
characteristic width $d$ of the inhomogeneity, in which case
$\tilde{g}(-2 q_x) \simeq \tilde{g}(0)$.
Using Eqs.~\eqref{eqn:graphene_conductivity} and \eqref{eqn:r_normal},
we find the reflection coefficient
\begin{equation}
	r \simeq i q_\infty \int\limits_{-\infty}^{\infty} d x 
	\left[
	\frac{\sigma(x)}{\sigma_\infty} - 1
	\right]
	\label{eqn:r_normal2}
\end{equation} 
for the normal incidence.
This simple equation gives a basic idea how $r$ may depend on $d$ and the local change in $\sigma$.

%%%%%%%%%%%%%%%%%%%%%%%%%%%%%%%%%%%%%%%%%%%%%%%%%%%%%%%%%%%%%%%%%%%%%
\section{Fitting the near-field profiles}
\label{sec:fitting}

As described in the main text, the near-field amplitude $\bar{s}(x)$
and phase $\phi(x)$ measured in our imaging experiments reveals the presence of interference fringes,
i.e., spatial modulations near the nanotube.
For example, $\sim 20\%$ variations of  $\bar{s}(x)$
are seen in Fig.~4(e) of the main text. 
Assuming these relative modulations should be of the order of the plasmon reflection coefficient $r$,
we can use Eq.~\eqref{eqn:r_normal2} to estimate the
perturbation of the conductivity caused by the nanotube.
Using the representative value of $\lambda_\infty \sim 200\unit{nm}$
at frequency $\omega = 890\unit{cm}^{-1}$ at which
the effective permittivity is equal to
\begin{equation}
	\kappa(\omega) =
	\frac{\epsilon_\mathrm{vacuum}(\omega)+\epsilon_\mathrm{SiO_2}(\omega)}
	{2} = 2.2\,,
\end{equation}
we find $\sigma_\infty \approx 5 i e^2 / h$
from Eq.~\eqref{eqn:q_p}.
Hence, we can reproduce $|r| \sim 0.2$ if we assume, for example,
that the reactive part of the conductivity $\mathrm{Im}\, \sigma$
is constant,
while the dissipative part is enhanced to about
$\mathrm{Re}\, \sigma \sim 3 e^2 / h$ over an interval of
width $d = 10\unit{nm}$ near the origin.
These numbers are generally consistent with the estimates in the main text.

\begin{figure}[t]
	\includegraphics[width=2.3in]{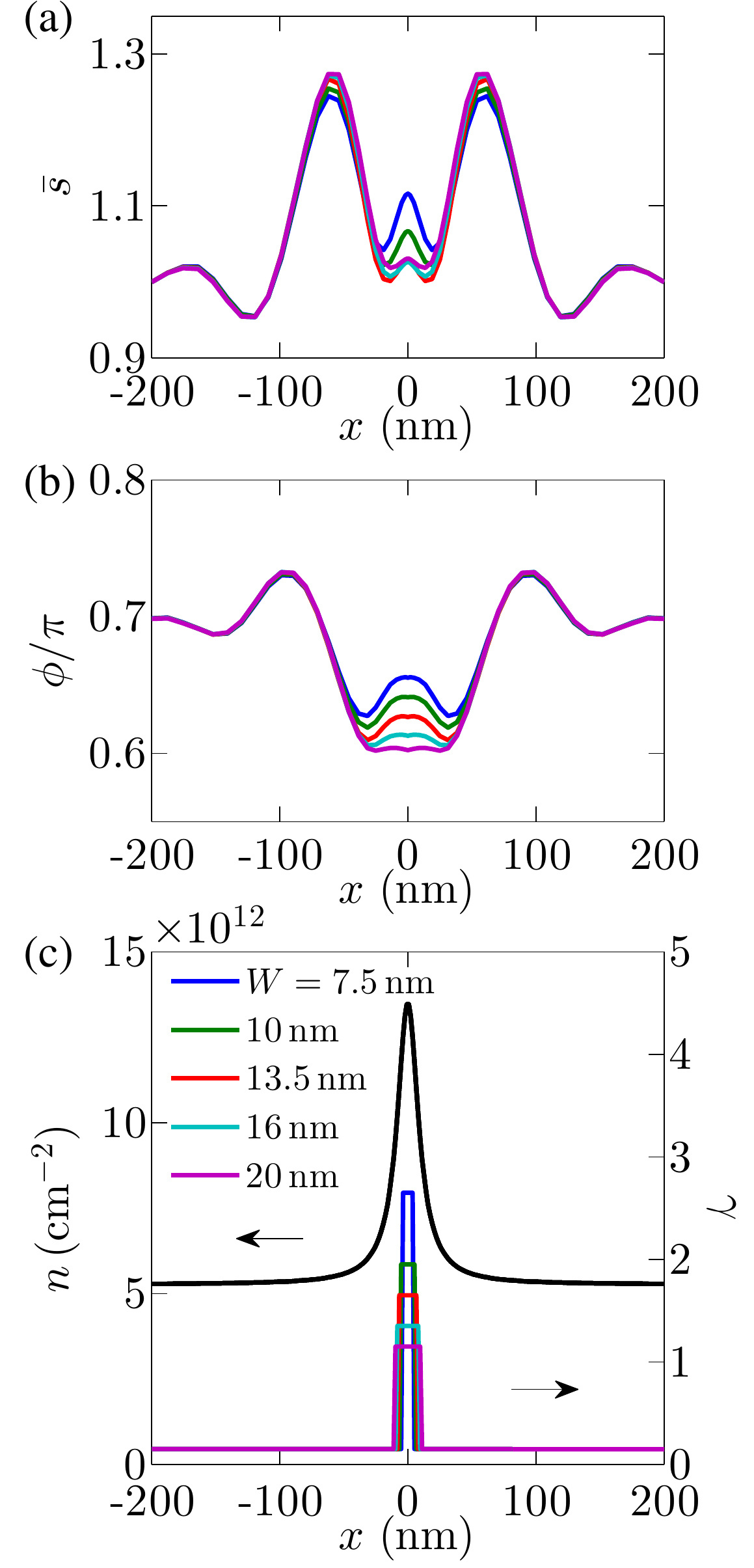}
	\caption{Simulated near-field amplitude $(\bar{s})$ and phase $(\phi)$ profiles for varying width $W$ of the box-like discontinuity in the damping rate $\gamma$.
		The height of the box is chosen so that the resultant near-field profiles are as close as possible to the one used in the actual fitting (red).
		The detailed shape of the box affects only small-distance features in the signal.
	}
	\label{fig:damp}
\end{figure}

To go beyond such rough estimates, additional modeling is required.
For it to be more realistic, several important issues need to be accounted for.
First, the plasmon waves launched and detected by the s-SNOM tip are not simple plane waves because the tip is positioned very close to the nanotube.
Second, the intensity of such waves depends in a nontrivial way on the electric field concentration that
occurs inside the tip-sample nanogap.
Third, in the experiment the tip-sample distance is varying periodically with the tapping frequency $\Omega$.
The complex near-field amplitude $s_3 e^{i \phi_3}$ corresponds to the signal demodulated at the third harmonic $3\Omega$.
The normalized signal $\bar{s}(x)$ is the ratio $s_3(x) / s_3(L)$,
where $L$ is a coordinate point giving a fair
approximation of the $x \to \infty$ limit.
[$L = 200\unit{nm}$ in Fig.~4(d)-(e) of the main text.]
Because of these complications, the quantitative modeling of $\bar{s}(x)$ and
$\phi(x)$ is possible only through numerical simulations.

Previously, an electromagnetic solver was developed,~\cite{Fei2012gtg, Fei2013epp} which takes these issues into account.
The algorithm implemented in the solver~\cite{Fei2013epp}
finds a numerical solution
of Eq.~\eqref{eqn:plasmon_eq}
discretized on a double grid of $q_y$ and $x$.
The external field is taken to be the sum of two terms.
The first one, a constant, represents the incident infrared beam.
The second one approximates the field created by the tip,
modeled as an elongated metallic spheroid.
The charge density distribution on the spheroid surface 
is found self-consistently from the condition that this surface is an equipotential.
The total dipole moment of the tip, which represents the instantaneous amplitude $s$ of the scattered electromagnetic field is computed.
Finally, $s_3 e^{i \phi_3}$ is calculated
by taking the appropriate Fourier transform and normalized to the reference point $x = L$ in order to yield
$\bar{s}$ and $\phi$.
The calculation is repeated for each tip position along the $x$-axis.

Using this solver we were able to produce simulated near-field profiles that matched well the measured ones
using a set of adjustable parameters.
We will now describe this fitting procedure and the results,
Figs.~\ref{fig:damp} and \ref{fig:parameters}.
As explained in Appendix~\ref{sec:r},
these fitting results should be considered an estimate of
the nonlocal conductivity of graphene averaged over the lengthscale $d$.

\begin{figure*}
	\includegraphics[width=2.3in]{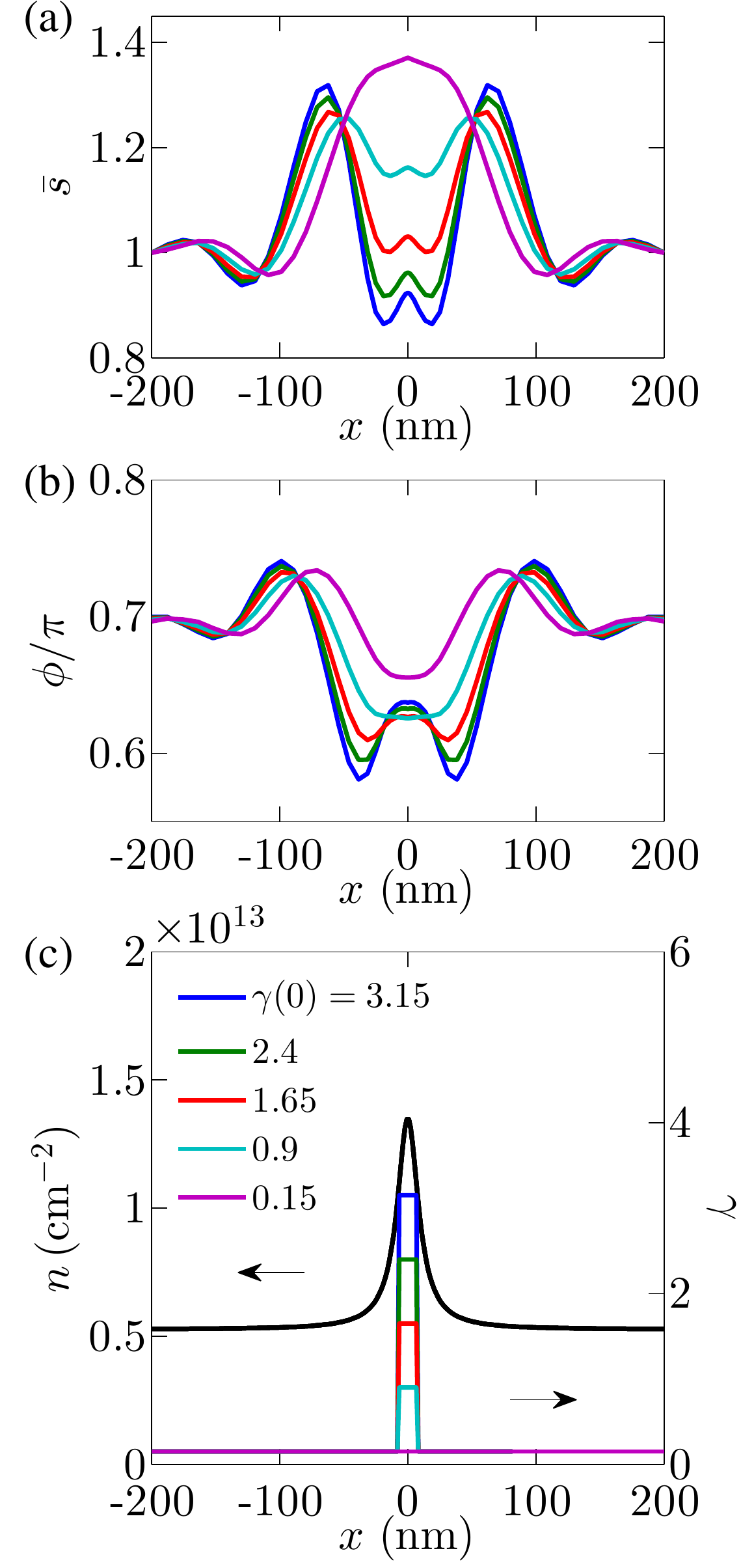}
	\includegraphics[width=2.3in]{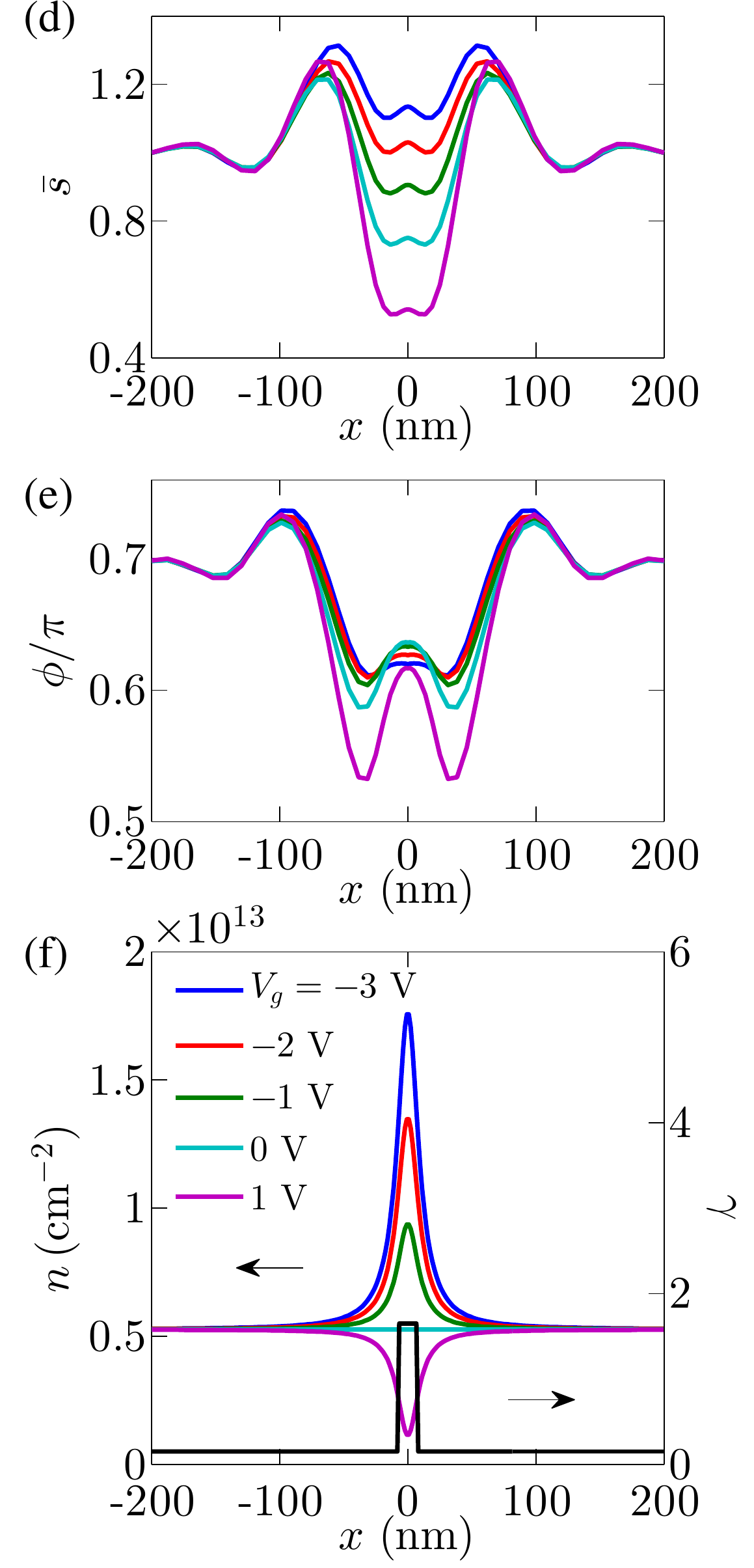}
	\includegraphics[width=2.3in]{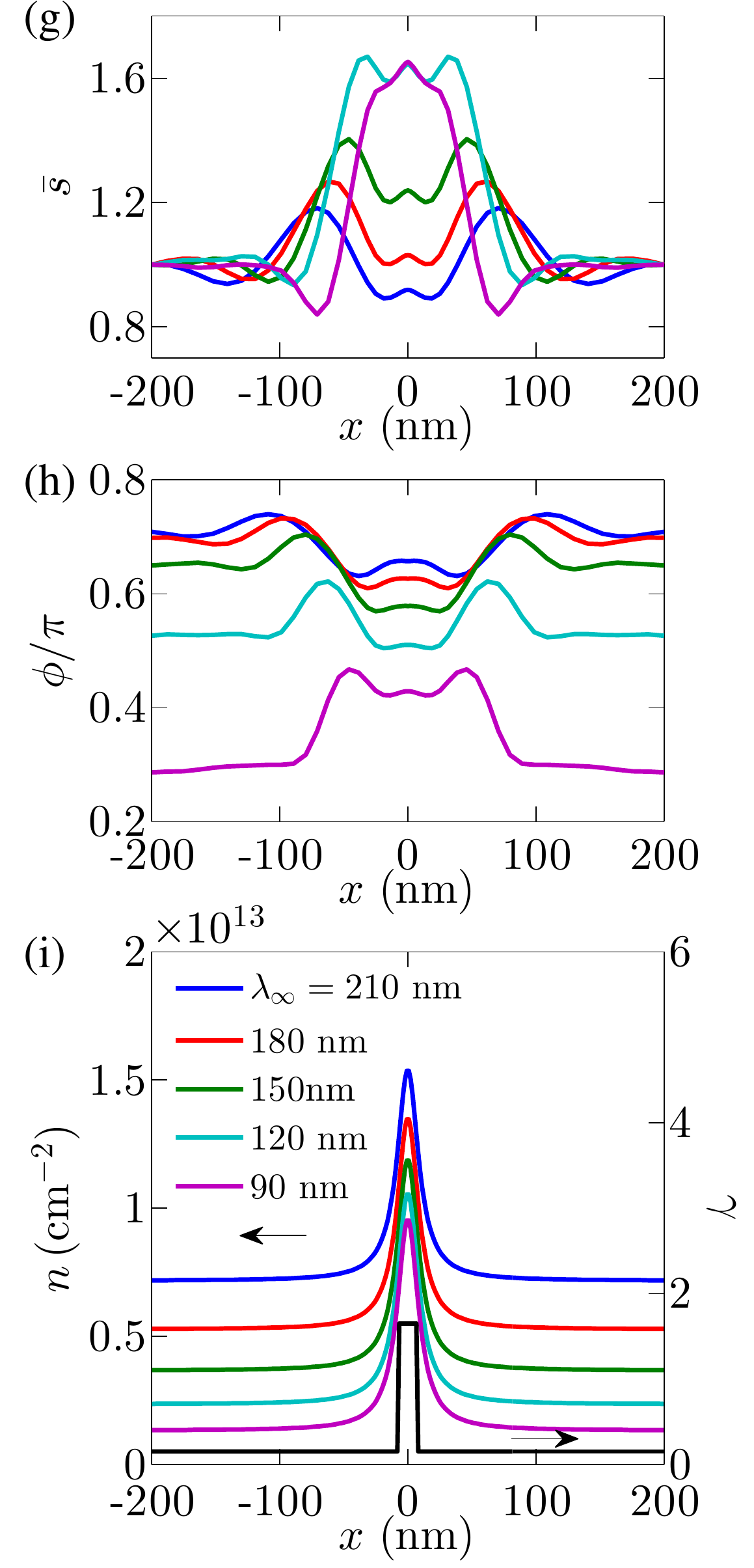}
	\caption{Simulated near-field amplitude $(\bar{s})$ and phase $(\phi)$ profiles along with the density $(n)$ and damping profiles $(\gamma)$ for (a)-(c) varying
		height $\gamma(0)$ of the box-like discontinuity  in the damping rate, (d)-(f) gate voltage $V_g$, or (g)-(i) background plasmon wavelength $\lambda_\infty$. 
		The red and black curves correspond to the profiles used to produce the fits in Fig.~4(e) of the main text.
	}
	\label{fig:parameters}
\end{figure*}

We took the trial damping function to be
\begin{equation}
	\gamma(x) = \gamma_\infty + [\gamma(0) - \gamma_\infty]\,
	\Theta\left(\frac{W}{2} - |x|\right)\,,
	\label{eqn:gamma_trial}
\end{equation}
where $\Theta(z)$ is the step-function,
see the colored boxes in Fig.~\ref{fig:damp}(c).
For the carrier density profile we assumed
the Lorentzian form [Fig.~\ref{fig:damp}(c), black curve]
\begin{equation}
	n(x) = n_\infty + \frac{C}{\pi}\,
	\frac{V_g}{e}\, \frac{d}{x^2+d^2}\,,
	\label{eqn:n}
\end{equation}  
where $d = 10\unit{nm}$ is the graphene-nanotube distance
and $C$ is the capacitance (per unit length) between them,
\begin{equation}
	C = \frac12 \, \frac{\kappa_0}{\ln ({2d}/{l})}\,.
\end{equation}
The effective static permittivity $\kappa_0$ of the dielectric environment around the nanotube is
\begin{equation}
	\kappa_0 =
	\frac{\epsilon_\mathrm{hBN}(0) + \epsilon_\mathrm{SiO_2}(0)}
	{2} = 3.7\,.
\end{equation}
Equation~\eqref{eqn:n} for $n(x)$ is appropriate for our relatively highly doped ($n > 10^{12}\unit{cm^{-2}}$) graphene which screens the electric field of the nanotube like a good metal.~\cite{Jiang2015erg}
We have not measured the radius $l \sim 1\unit{nm}$ of the nanotube
directly, so there is an uncertainty in $C$. 
This uncertainty is however small due to the logarithmic form of $C$. 
On the other hand, the voltage difference $V_g$ between the nanotube and graphene is measured.
Hence, our model contains four adjustable parameters: $\gamma(0)$, $\gamma_\infty$, $W$, and $n_\infty$. 
Instead of the last of these we can use the asymptotic plasmon wavelength $\lambda_\infty$ because they are directly related via Eqs.~\eqref{eqn:graphene_conductivity}, \eqref{eqn:q_p},
\eqref{eqn:lambda}, and one more equation,
\begin{equation}
\mu(n) = \hbar v_F (\pi |n|)^{1 / 2}\,.
\label{eqn:mu_n}
\end{equation}

A brief comment on the trial form of $\gamma$ and $n$ may be in order.
The discontinuous box-like profile of the dimensionless damping rate $\gamma(x)$ may seem artificial; however,
since the plasmon wavelength is much larger than the width of the box $W\sim d$, the near-field amplitude
is largely insensitive to the precise functional form of $\gamma(x)$.
In principle, we could also choose a box-like profile for $n(x)$.
However, Eq.~\eqref{eqn:n} is just as convenient and is better physically motivated.

In Fig.~\ref{fig:damp}(a, b) we show the simulated profiles
of the near-field amplitude $\bar{s}$ and phase $\phi$
for several $W$'s for fixed $V_g = -2\unit{V}$, $\lambda_\infty = 180\unit{nm}$, and $\gamma(\infty) = 0.15$. 
The profiles for $W = 13.5 \unit{nm}$ and $\gamma(0) = 1.65$ match those measured in the experiment rather well.
This fitting suggests that the electrified nanotube
causes the density change by almost a factor of three and the damping enhancement by more than an order of magnitude.
The former conclusion should be robust as it is a consequence of the simple electrostatics. On the other hand, the latter
should be considered the experimental discovery.
An explanation for this surprisingly high local damping was presented in the main text and
the technical details were given in Appendix~\ref{sec:square_well}.

A rough correspondence between the numbers obtained above and the parameters of the square-well model can be established as follows.
The depth $u$ of the well is taken as the integrated potential divided by the width $d$ of the well, $u=\frac1d\int v(x) dx$. 
The potential $v(x)$ can be found through $v(x)=\mu(x)-\mu_\infty$ and Eq.~\eqref{eqn:mu_n}.
This results in the dimensionless well depth $U=ud/\hbar v_F=13$.   
The estimation of the integrated conductivity $\bar{\sigma}$ is more complicated.
Due to the presence of a nonzero background $\gamma_\infty$, density variations will also contribute to the real part of the optical conductivity.
To isolate the contribution of the optical transitions, the integrated conductivity is calculated as $\bar{\sigma}=\frac1d\,\mathrm{Re}\int[\sigma(x)-\sigma'(x)] dx$, where $\sigma'(x)$ is the conductivity of a comparison system, which has the same density profile $n(x)$ but the constant damping rate $\gamma=\gamma_\infty$.
This prescription yields $\bar{\sigma}=3.5 \unit{e^2/h}$.

The dependence of $\bar{s}$ and $\phi$ profiles on the other adjustable parameters, such as $\gamma(0)$, $\lambda_\infty$,
and also on the gate voltage $V_g$
is illustrated in Fig.~\ref{fig:parameters}(a, b), 
(g, h), and (d, e), respectively.
The profiles vary dramatically with the changes in these parameters, and so
the determination of the best-fitting values of the adjustable parameters
has very little uncertainty.
This analysis is yet another illustration of how the s-SNOM nanoimaging can be a powerful and sensitive technique for probing the local surface conductivity of graphene and perhaps many other 2D materials as well.

%%%%%%%%%%%%%%%%%%%%%%%%%%%%%%%%%%%%%%%%%%%%%%%%%%%%%%%%%%%%%%%%%%%%%%%%%%%%%
\section{Device Fabrication}
\label{sec:fabrication}

\begin{figure}[htb]
\includegraphics[width=3.4 in]{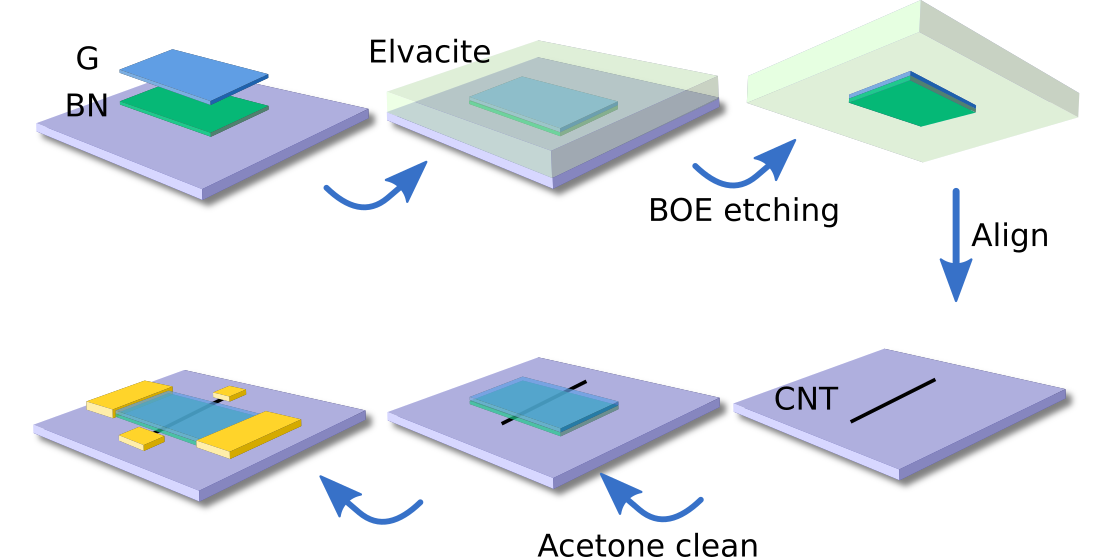}
\\[24pt]
\includegraphics[width=3.2 in]{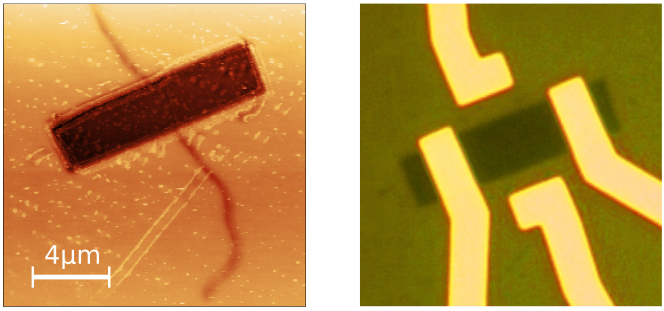}
\caption{(Top) Fabrication process of the graphene/hBN/CNT heterostructure on a SiO$_2$/Si substrate. 
(Bottom left) AFM image of the device before addition of the contacts. (Bottom right) Optical image of the completed device.
}
\label{fig:fab}
\end{figure}

Our device consists of (from top to bottom)
a graphene monolayer, a $10\unit{nm}$-thick hBN flake, a metallic single-wall CNT, and a SiO$_2$/Si substrate.
The CNT was grown by chemical vapor deposition, located using a scanning electron microscope, and transferred onto a SiO$_2$/Si substrate.
Monolayer graphene was mechanically exfoliated.
Using a PPC/PDMS stamp, the graphene/hBN stack was transferred onto a separate SiO$_2$/Si substrate.
The stack was then picked up with an acrylic resin Elvacite~\cite{Huang2015scc}.
It was subject to buffered oxide etch (BOE), aligned, and transferred to cover the CNT.
The Elvacite was cleaned away with acetone. 
The final step of the fabrication was adding metallic contacts to the CNT and graphene. 
These steps are summarized in Fig.~\ref{fig:fab} (top).
The AFM and optical images of the device are shown in
Fig.~\ref{fig:fab} (bottom left and right). 

%%%%%%%%%%%%%%%%%%%%%%%%%%%%%%%%%%%%%%%%%%%%%%%%%%%%%%%%%%%%%%%%%%%%%%%%%%%%%
\section{Supercritical transitions}
\label{sec:supercritical}

\begin{figure}[b]
\includegraphics[width=3.4 in]{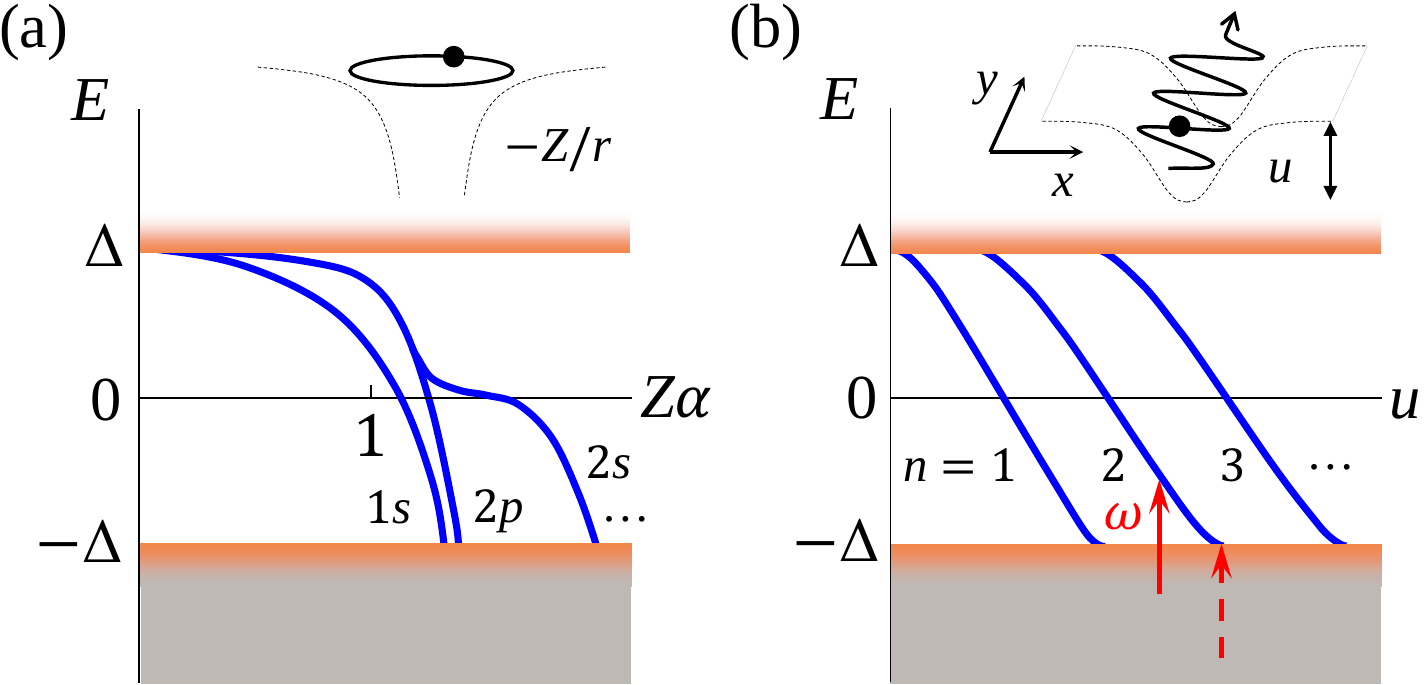}
\caption{(Color online) Bound-states energies (blue) of Dirac fermions in strong external potentials.
The orange lines represent the boundaries of continuum states.
The insets depict classical trajectories (solid) and the
potential profiles (dashed).
(a) Low-lying states of large-$Z$ atoms. [Adopted from Ref.~\onlinecite{Zeldovich1972ess}.]
(b) Fixed-$k_y$ states in a 1D potential well in graphene.
The solid red arrow shows an optical transition of frequency $\omega$ from the valence band to a bound state.
This transition disappears at some critical well depth $u$ (dashed arrow).
}
\label{fig:atomic_spec}
\end{figure}

Our problem has an intriguing parallel to the collapse of superheavy atoms
in nuclear physics, which is as follows.
For a very large nuclear charge $Z > Z_c \sim 1 / \alpha$, the extrapolated
values of the first few atomic levels sink below
the top of the positron energy band,~\cite{Zeldovich1972ess} $-\Delta = -m_0 c^2$,
see Fig.~\ref{fig:atomic_spec}(a).
(Here $\alpha = e^2 / \hbar c$ is the fine-structure constant and $m_0$ is the electron mass.)
Such supercritical states can no longer be bound to the nucleus but should be quasi-bound, being hybridized with the extended states in the positron band.
In graphene where the role of $c$ is played by the Fermi velocity $v_F\sim c / 300$, the critical charge
is rather small, $Z_c\sim 1$~\cite{Shytov2007vps, Pereira2007cip}.
This has made it possible to observe the long-sought supercriticality
experimentally by measuring the tunneling density of states
near charged impurities~\cite{Wang2013oac}.
Analogous transitions~\cite{Kennedy2002wsp} are possible for the bound states studied in this Letter, Fig.~\ref{fig:atomic_spec}(b).
Compared to prior studies of a single~\cite{Shytov2007vps, Pereira2007cip, Fogler2007shc} or a few pointlike charges~\cite{Wang2013oac, Pereira2007cip, DeMartino2014edi, Gorbar2015snt}, the 1D geometry 
examined in this work 
has several advantages.
The gapless 2D Dirac spectrum is replaced by a gapped one with $\Delta = |\hbar k_y v_F|$ [Fig.~\ref{fig:atomic_spec}(b)],
making the analogy to the atomic collapse problem~\cite{Zeldovich1972ess} closer.
Alternatively, it prompts an analogy to a hypothetical
cosmic string~\cite{Nascimento1999cvc},
previously used in graphene literature~\cite{Pereira2010gme, Chakraborty2013tcs} in a different context.
More importantly, our ``1D atom'' permits continuous \textit{in situ} tunability in terms of at least two
parameters: the gate voltage and the optical excitation frequency.
Experimental verification of these supercritical transitions can be attempted via two complementary approaches.
One is to examine the changes in the LDOS.
In fact, the detailed calculations of the LDOS presented in Appendix~\ref{sec:LDOS} were done exactly with such experiments in mind.
The other approach is to look for abrupt drops in the local optical conductivity $\bar{\sigma}$ caused by the liquidation of the
optical transitions, see Fig.~3(b) of the main text. 
Unfortunately, in either the conductivity or the LDOS, the supercritical signatures are very subtle compared to those of, say, van Hove singularities. Also, pinpointing these transitions requires an exhaustive search of the parameter space, which, for technical reasons, has not been possible in the devices we fabricated so far. Nevertheless, this can be an interesting and challenging problem for future work.

\bibliography{line_charge_plasmon}
\end{document}